\RequirePackage{fix-cm}
\documentclass[twocolumn,epjc3-upd,fleqn]{svjour3}
\usepackage{newtxtext,newtxmath}
\RequirePackage{graphicx}
\RequirePackage{subfigure}
\RequirePackage{rotating}
\RequirePackage{pdflscape}
\RequirePackage{color}
\usepackage{mathtools}
\RequirePackage{latexsym}
\RequirePackage{xspace}
\RequirePackage[numbers,sort&compress]{natbib}
\RequirePackage[colorlinks,citecolor=blue,urlcolor=blue,linkcolor=blue]{hyperref}
\usepackage{url}

\usepackage{amsmath,amsfonts,bm}
\usepackage{wasysym}
\usepackage{color}
\usepackage{lineno}
\usepackage{multirow}
\usepackage[usenames,dvipsnames,svgnames,table]{xcolor}	

\let\oldequation\equation
\let\oldendequation\endequation

\renewenvironment{equation}
  {\linenomathNonumbers\oldequation}
  {\oldendequation\endlinenomath}

\let\oldalign\align
\let\oldendalign\endalign

\renewenvironment{align}
  {\linenomathNonumbers\oldalign}
  {\oldendalign\endlinenomath}

\usepackage{acronym}
\usepackage[normalem]{ulem}

\usepackage{lineno}

\newcommand{\onbb}{$0\nu\beta\beta$\xspace}

\newcommand{\Qbb}{$Q_{\beta\beta}$\xspace}

\journalname{Eur. Phys. J. C}

\begin{document}

\title{Sensitivity of the CUPID experiment to $0\nu\beta\beta$ decay of $^{100}$Mo}
\author{K.~Alfonso\thanksref{VT_US,UCLA_US}
\and
A.~Armatol\thanksref{LBNL_US,fn0}
\and
C.~Augier\thanksref{IP2I_France}
\and
F.~T.~Avignone~III\thanksref{UofSC_US}
\and
O.~Azzolini\thanksref{LNL_Italy}
\and
A.S.~Barabash\thanksref{NRC_KI_Russia}
\and
G.~Bari\thanksref{SdB_Italy}
\and
A.~Barresi\thanksref{UniMIB_Italy,MIB_Italy}
\and
D.~Baudin\thanksref{CEA_IRFU_France}
\and
F.~Bellini\thanksref{SdR_Italy,SURome_Italy}
% ...
}

\institute{Virginia Polytechnic Institute and State University, Blacksburg, VA, USA\label{VT_US}
\and
University of California, Los Angeles, CA, USA\label{UCLA_US}
\and
Lawrence Berkeley National Laboratory, Berkeley, CA, USA\label{LBNL_US}
\and
Univ Lyon, Universit\'e Lyon 1, CNRS/IN2P3, IP2I-Lyon, Villeurbanne, France\label{IP2I_France}
\and
University of South Carolina, Columbia, SC, USA\label{UofSC_US}
\and
INFN Laboratori Nazionali di Legnaro, Legnaro, Italy\label{LNL_Italy}
\and
National Research Centre Kurchatov Institute, Kurchatov Complex of Theoretical and Experimental Physics, Moscow, Russia\label{NRC_KI_Russia}
\and
INFN Sezione di Bologna, Bologna, Italy\label{SdB_Italy}
\and
Universit\`a degli Studi di Milano-Bicocca, Dipartimento di Fisica, Milano, Italy\label{UniMIB_Italy}
\and
INFN Sezione di Milano-Bicocca, Milano, Italy\label{MIB_Italy}
\and
IRFU, CEA, Universit\'e Paris-Saclay, Saclay, France\label{CEA_IRFU_France}
\and
INFN Sezione di Roma, Rome, Italy\label{SdR_Italy}
\and
Sapienza University of Rome, Rome, Italy\label{SURome_Italy}
% ...
\and
INFN Laboratori Nazionali del Gran Sasso, Assergi (AQ), Italy\label{LNGS_Italy}
\and
INFN Laboratori Nazionali di Frascati, Frascati, Italy\label{LNF_Italy}
\and
Universit\'e Paris-Saclay, CNRS/IN2P3, IJCLab, Orsay, France\label{IJCLab_France}
\and
CNR-Institute for Microelectronics and Microsystems, Bologna, Italy\label{CNR-IMM_Italy}
\and
INFN Sezione di Pavia, Pavia, Italy\label{PV_Italy}
% ...
\and
University of California, Berkeley, Berkeley, CA, USA\label{UCB_US}
\and
INFN Sezione di Genova, Genova, Italy\label{SdG_Italy}
% ...
\and
Rome Tor Vergata University, Rome, Italy\label{Rome_Tor_Vergata_University_Italy}
\and
INFN sezione di Roma Tor Vergata, Rome, Italy\label{INFN_Tor_Vergata_Italy}
\and
Northwestern University, Evanston, IL, USA\label{NWU_US}
\and
Argonne National Laboratory, Argonne, IL, USA\label{ANL_US}
% ...
\and
Massachusetts Institute of Technology, Cambridge, MA, USA\label{MIT_US}
\and
Boston University, Boston, MA, USA\label{BU_US}
\and
California Polytechnic State University, San Luis Obispo, CA, USA\label{CalPoly_US}
% ...
\and
Yale University, New Haven, CT, USA\label{Yale_US}
\and
Department of Physics and Astronomy, University of Pittsburgh, Pittsburgh, PA, USA\label{Pittsburgh_US}
\and
Drexel University, Philadelphia, PA, USA\label{Drexel_US}
\and
Johns Hopkins University, Baltimore, MD, USA\label{JHU_US}
% ...
\and
Nikolaev Institute of Inorganic Chemistry, Novosibirsk, Russia\label{NIIC_Russia}
\and
INFN Sezione di Padova, Padova, Italy\label{PD_Italy}
\and
Univ. Grenoble Alpes, CNRS, Grenoble INP, SIMAP, Grenoble, France\label{SIMaP_Grenoble_France}
\and
University of Bologna, Bologna, Italy\label{UnivBologna_Italy}
}

\thankstext{fn0}{Now at IP2I-Lyon, Univ Lyon, France}
\thankstext{fn1}{Now at Physik-Institut, University of Z\"urich, Z\"urich, Switzerland}
\thankstext{fn2}{Now at University College London, London, UK}
\thankstext{fn3}{Also at Instituto de F{\'\i}sica, Universidade de S\~ao Paulo, Brazil and Max-Planck-Institut f\"ur Physik, M\"unchen, Germany}

\date{Received: date / Accepted: date}

\maketitle

\begin{abstract}
CUPID is a next-generation bolometric experiment to search for neutrinoless double-beta decay ($0\nu\beta\beta$) of $^{100}$Mo using Li$_2$MoO$_4$ scintillating crystals. It will operate 1596 crystals at $\sim$10 mK in the CUORE cryostat at the Laboratori Nazionali del Gran Sasso in Italy. Each crystal will be facing two Ge-based bolometric light detectors for $\alpha$ rejection. We compute the discovery and the exclusion sensitivity of CUPID to $0\nu\beta\beta$ in a Frequentist and a Bayesian framework. This computation is done numerically based on pseudo-experiments. For the CUPID \emph{baseline} scenario, with a background and an energy resolution of $1.0 \times 10^{-4}$ counts/keV/kg/yr and  5 keV FWHM at the Q-value, respectively, this results in a Bayesian exclusion sensitivity (90\% c.i.) of $\hat{T}_{1/2} > 1.6 \times 10^{27} \ \mathrm{yr}$, corresponding to the effective Majorana neutrino mass of  $\hat{m}_{\beta\beta} < \ 9.6$ -- $28 \ \mathrm{meV}$. 
The Frequentist discovery sensitivity (3$\sigma$) is $\hat{T}_{1/2}= 1.0 \times 10^{27} \ \mathrm{yr}$, corresponding to $\hat{m}_{\beta\beta}= \ 12$ -- $36 \ \mathrm{meV}$.
\end{abstract}
\keywords{Double-beta decay, bolometers, scintillating crystals and particle identification}
\sloppy
\section{Introduction}

Double-beta decay with neutrino emission ($2\nu\beta\beta$) is an extremely rare process in which two neutrons of a nuclei transform into two protons emitting two electrons and two antineutrinos. Such a process is allowed by the Standard Model of particle physics (SM) and has been observed for eleven isotopes with half-lives in the range of $10^{18}$--$10^{24}$ yr \cite{Barabash:2b2n,Pritychenko:2024ase}. The neutrinoless double-beta decay ($0\nu\beta\beta$) is a hypothetical process in which two neutrons transform into two protons emitting two electrons with no corresponding emission of antineutrinos.  The observation of $0\nu\beta\beta$ would open the door to physics beyond the SM \cite{Agostini:2022zub,DellOro:2016tmg,Dolinski:2019nrj,Gomez-Cadenas:2023vca,Bossio:2023wpj} because it would directly imply lepton number violation. Moreover, the detection of \onbb would demonstrate that the neutrino is the only fermion to be a Majorana particle, showing that it is its own anti-particle. The existence of a Majorana neutrino would also have crucial consequences in cosmology, as it could have a role in the baryon asymmetry in the Universe through leptogenesis \cite{Fukugita:1986hr,Deppisch:2017ecm}. 
\onbb decay may be induced by a plethora of mechanisms, however,  a prevalent interpretation is a minimal extension of the SM which includes the exchange of a light Majorana neutrino. In this scenario,  the $0\nu\beta\beta$ decay rate is proportional to the square of the effective Majorana neutrino mass, $m_{\beta\beta}$$^2$, with  $m_{\beta\beta}$ related to the absolute mass of the neutrinos \cite{ParticleDataGroup:2024cfk}. Therefore, the observation of the $0\nu\beta\beta$ provides a way to constrain neutrino masses.

The minimal $0\nu\beta\beta$ decay signal is a peak in the summed electron energy spectrum, positioned exactly at the Q-value of the process (Q$_{\beta\beta}$). This specific signal allows to discriminate the $0\nu\beta\beta$ from $2\nu\beta\beta$, which produces a continuous spectrum.
 
A very sensitive approach for discovering $0\nu\beta\beta$ consists of a detector that incorporates the source isotope, ensuring a high detection efficiency. Furthermore, the detector should exhibit a good energy resolution and low background, as well as a capability of a long-term stable operation.

The bolometric technology fulfils all the required features.  Bolometers can embed the nuclei of interest, they have excellent energy resolution $\mathcal{O}$ (1--10 keV) and they can be operated in low background conditions. Cryogenic bolometers are crystals operated at temperatures of $\sim$10~mK. A particle releasing its energy in the crystal induces a rise in the temperature, which is read by a thermal sensor. The strength of this technique for $0\nu\beta\beta$ searches is demonstrated by the success of the CUORE (Cryogenic Underground Observatory for Rare Events) experiment \cite{CUORE:2002myo,CUORE:2015hsf,CUORE:2021ctv,CUORE_nature, CUORE:2024ikf}. CUORE, located at the Laboratori Nazionali del Gran Sasso (LNGS), collected data with more than two ton-years of exposure so far, showing the feasibility of operating $\sim$1000 crystals at $\sim$10 mK over several years.
CUORE searches for the $0\nu\beta\beta$ of $^{130}$Te with TeO$_2$ crystals and has set the best limit on the $^{130}$Te half-life \cite{CUORE:2024ikf}. However, the CUORE sensitivity is currently limited by a background induced by $\alpha$ particles originating from surface contaminations of the detector structure. This motivated the development of scintillating bolometers \cite{Poda:2021hsv}. Thanks to a dual read-out of heat and scintillation light signals, $\alpha$ particles can be rejected due to their different light yield with respect to $\beta/\gamma$ particles. Such detectors were developed and tested through the CUPID-0 and CUPID-Mo demonstrators \cite{CUPID:2018kff, Armengaud:2019loe}. Both experiments have shown the ability to discriminate $\alpha$ particles at a level higher than $99.9 \%$ \cite{CUPID:2019lzs,Armengaud:2019loe} and have set leading  $0\nu\beta\beta$ limits for $^{82}$Se \cite{CUPID:2022puj} and $^{100}$Mo \cite{Augier:2022znx}, respectively. The next-generation bolometric $0\nu\beta\beta$ experiment CUPID (CUORE Upgrade with Particle IDentification) will search for the $0\nu\beta\beta$ of $^{100}$Mo embedded in scintillating Li$_2$MoO$_4$ crystals \cite{CUPID_Baseline,CUPID_CDR}. The background is expected to be reduced by a factor $100$ with respect to CUORE ($\sim$10$^{-2}$ counts/keV/kg/yr), primarily thanks to the $\alpha$ rejection and to the higher Q-value of  $^{100}$Mo ($Q_{\beta\beta}$ = 3034 keV) implying a signal located outside the bulk of natural radioactivity. CUPID aims to fully explore the $0\nu\beta\beta$ in the inverted ordering (IO) regime of neutrino masses \cite{Capozzi:2025wyn,ParticleDataGroup:2024cfk}, as well as the normal ordering (NO) regime for a neutrino mass larger than 10~meV.

In this paper, we develop a statistical analysis to compute the discovery and exclusion sensitivity of CUPID to the $0\nu\beta\beta$ half-life and to the effective Majorana neutrino mass. We present briefly in Section \ref{sec:cupid} the CUPID design and the expected detector performances. In Section \ref{sec:stat} we describe the general statistical methodology  employed in this work. Section \ref{sec:freq} details the Frequentist analysis we developed to estimate the discovery and the exclusion sensitivity. It provides the resulting sensitivities on half-life of $0\nu\beta\beta$ and $m_{\beta\beta}$ assuming the CUPID \emph{baseline} scenario. We also discuss the impact  of varying the background and the energy resolution on the sensitivity. 
Section \ref{sec:Bay} presents the methodology used in the framework of a Bayesian analysis and the resulting  exclusion sensitivities on the $0\nu\beta\beta$ half-life  and $m_{\beta\beta}$, evaluated for various background indices and energy resolutions.

\section{The CUPID experiment}
\label{sec:cupid}

The CUPID experiment \cite{CUPID_Baseline, CUPID_CDR} will be located at LNGS and installed in the CUORE cryostat. The {\it baseline} CUPID detector design consists of 1596 cubic scintillating Li$_2$MoO$_4$  crystals, enriched to $\geq$ 95\% in $^{100}$Mo. The size of an individual crystal is $45 \times 45 \times 45$ mm$^3$, corresponding to a mass of $\sim$ 280~g each, for a total mass of 450 kg or a  $^{100}$Mo mass of 240~kg. Each crystal will be facing two Ge-wafer light detectors \cite{CUPID:2022opf}. A Neutron Transmutation Doped Ge thermistor (NTD) \cite{10.1117/12.176771} will be glued to the crystals and the light detector to measure the thermal and scintillation light signals. To improve rejection capabilities, the light detectors will be instrumented with concentric Al electrodes to exploit the Neganov-Trofimov-Luke (NTL) amplification mechanism \cite{Luke:1988, Neganov:1985khw, NOVATI2019320, Ahmine:2023xhg, Armatol:2025zrq}. The detectors will be operated at a temperature around 10–20 mK.


The sensitivity of CUPID depends on several parameters of the detector performance, i.e. isotope enrichment, containment and selection efficiencies, background, energy resolution, crystal mass and livetime. Data from demonstrators, the  pre-procurement of crystals and the operational experience of CUORE give confidence that most of these parameters are  well known with $\mathcal{O}$(1\%) level uncertainties \cite{Augier:2022znx,CUPID-Mo:2023lru}. However, the background index ($B$) as well as the detector resolution are subject to a larger uncertainty as improvements in the design of the detector structure of CUPID and a new NTD optimization and irradiation campaign still need to be fully evaluated. 
Table \ref{tab:perf} summarises the parameters of the CUPID {\it baseline} design.
The projected background index is evaluated in a 30 keV region of interest (ROI) around $Q_{\beta\beta}$, corresponding to six times the target energy resolution. Background sources polluting the ROI include environmental radiation (muons, neutrons), intrinsic radioactivity ($^{214}$Bi and $^{208}$Tl $\beta$ decays), and pile-up from coincident events in a single crystal. Material radioactivity estimates are derived from radiopurities obtained in the CUORE \cite{CUORE:2024fak} and CUPID-Mo \cite{CUPID-Mo:2023vle} models, which are used as an input in our GEANT4 simulations. This results in  $B = 6.0 \times 10^{-5}$ counts/keV/kg/yr, dominated by nearby components ($4.0 \times 10^{-5}$ counts/keV/kg/yr). Pile-ups are mainly coming from $^{100}$Mo $2\nu\beta\beta$ decays, which produce 2.6 mHz rate per crystal. Its suppression depends on detector time resolution and signal-to-noise ratio. Implementing NTL amplification in light detectors is expected to lower this contribution below $5.0 \times 10^{-5}$ counts/keV/kg/yr \cite{Ahmine:2023xhg, Armatol:2025zrq}. Simulations predict $\sim10$ muons/h in the CUPID setup \cite{CUORE:2025lzd}. We aim for a $1.0 \times 10^{-6}$ counts/keV/kg/yr muon veto with a composite system out of plastic scintillator \cite{Moore:2025eil} and a water Cherenkov detector. Together with the addition of 10 cm of polyethene to the already existing 20 cm layer in the lateral and the bottom parts of the CUORE cryostat, simulation studies show that the neutron-induced $B$ is reduced to $ \sim 2 \times 10^{-6}$ counts/keV/kg/yr. These studies result in a total projected $B\sim$$1 \times 10^{-4}$ counts/keV/kg/yr \cite{CUPID_Baseline, loaiza:tel-04633827, imbert:tel-04266831}.
We expect that the novel detector structure, which is designed to be machined through contactless techniques like laser cutting and with  simpler shapes compared to CUORE, will result in easier surface cleaning and improved surface radiopurity. Together with improvements in the cleaning procedure and detector technology, we hence have significant potential for a further reduction of the background index.
Thus, in our sensitivity study, we considered as well a  $B = 0.6 \times 10^{-4}$ counts/keV/kg/yr. We also evaluated a best case scenario under the assumption of suppressing backgrounds related to close contaminations in the towers and  2$\nu\beta\beta$ pile-up to sub-dominant
levels, which corresponds to $0.2 \times 10^{-4}$ counts/keV/kg/yr. We also considered an ultra-conservative scenario with a 
$B = 1.5 \times 10^{-4}$ counts/keV/kg/yr. 

The detector energy resolution depends on many parameters linked to the electro-thermal properties of the bolometers, including the crystal itself, the thermal sensor (for example, the sensor dimensions, the coupling to the crystals), the vibrational noise in the cryogenic set-up and the electronics. 
The interplay of these parameters is complex and hence difficult to anticipate for the final towers mounted in the cryostat.  The projected energy resolution at Q$_{\beta\beta}$ for CUPID is 5 keV (FWHM), based on first tests of a CUPID \emph{baseline} module \cite{CUPID:2022opf}. More recently, in a first test of a full CUPID tower in LNGS, we demonstrated an energy resolution of ($6.2 \pm 1.9$) keV FWHM at 2615 keV \cite{CUPID:2025wbt}. Improved resolution is expected in the  CUORE cryostat, which will be upgraded to minimize vibrations and equipped with diagnostic devices to allow for monitoring and for denoising in the analysis \cite{Vetter:2023fas}. Thermal sensors' dimensions and coupling to the crystals will also be optimized. We studied the impact on the discovery sensitivity of an energy resolution of 7.5 keV (as the one in the CUPID-Mo demonstrator installed in the cryostat of the EDELWEISS dark matter experiment, not optimized for \onbb searches \cite{Armengaud:2019loe}) and a very extreme case with an energy resolution of 10 keV.

\begin{table}[]
    \centering
    \caption{Parameters of the CUPID \emph{baseline} design.}
    \begin{tabular}{c|c}
        Parameter & Value \\ \hline \hline
   Crystal Mass       &  $450$ kg \\
   Isotope enrichment & $95$\% \\
   Energy resolution & $5$ keV (FWHM) \\
   Livetime & 10 years \\
   Containment efficiency & 79\% \\
   Selection efficiency & 90\% \\
   Background index & $10^{-4}$ counts/keV/kg/yr \\
    \end{tabular}
    \label{tab:perf}
\end{table}

\section{Statistical methodology}
\label{sec:stat}

\begin{figure}
    \centering
    \includegraphics[width=0.45\textwidth]{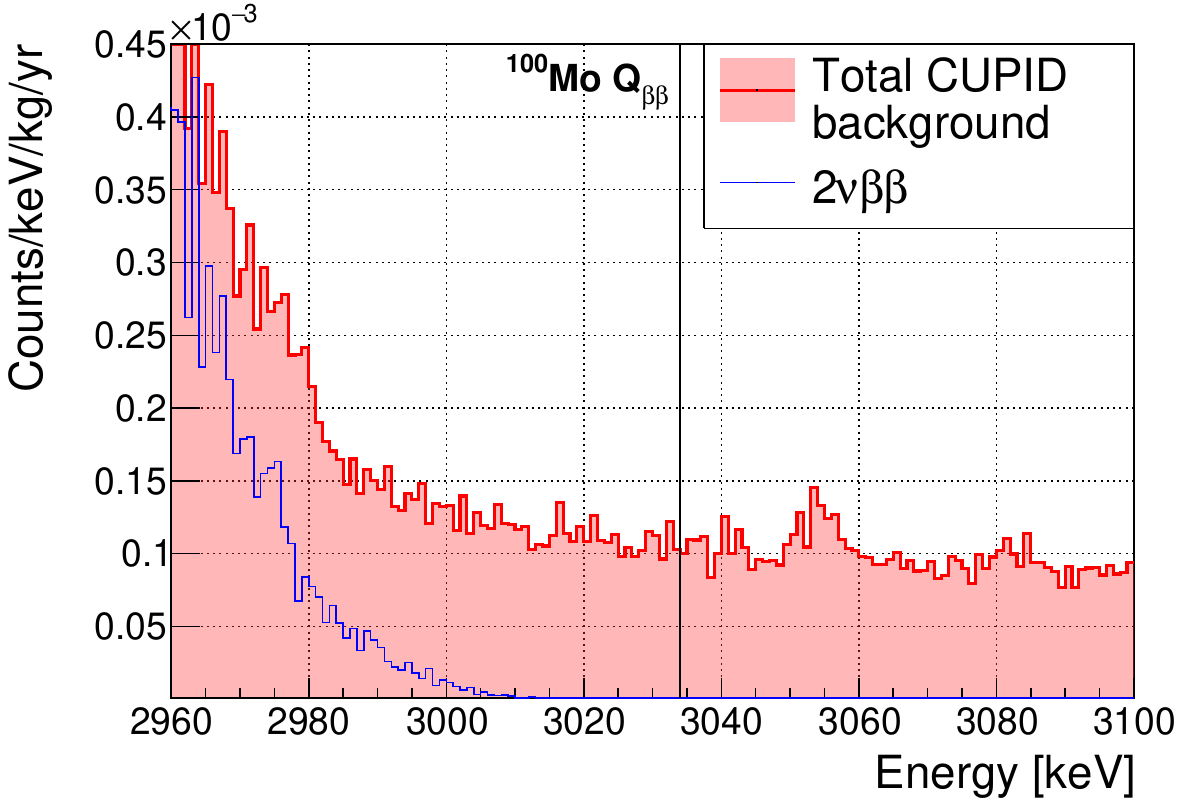}
    \caption{Expected total background in the region of interest of CUPID, in red, including the $2\nu\beta\beta$ highlighted in blue. The histograms are obtained based on GEANT4 simulations.}
    \label{fig:cupid_bkg}
\end{figure}

Our statistical analysis, both Frequentist and Bayesian, starts from a likelihood-based framework. Given a model of the data with a parameter or set of parameters $\Theta$ and a dataset $\mathcal{D}$, the likelihood is defined as:
\begin{equation}
    \mathcal{L} = p(\mathcal{D}|\Theta),
\end{equation}
or the probability of observing the data $\mathcal{D}$ given the value of the set of parameters $\Theta$. In our case, the experimental data consist of a vector of the energies of observed events passing all selection cuts, $\vec{x}$. Our model is parameterised by the $0\nu\beta\beta$ decay rate given by the inverse of the half-life, $\Gamma =1/T_{1/2}^{0\nu}$, and a set of nuisance parameters $\vec{\nu}$, for example, those describing the background rate. We write the model as  $f(x_i;\Gamma,\vec{\nu})$. The likelihood is given by an extended unbinned likelihood \cite{Cowan}:
\begin{equation}
\label{likelihood}
    \mathcal{L}=\text{Pois}\big(N;\lambda\big)\prod_{i=1}^N f(x_i;\Gamma,\vec{\nu}),
\end{equation}
where the number of counts (number of elements in the vector $\vec{x}$), $N$, is itself a random variable that follows a Poisson distribution with a mean  $\lambda$.
Our model of the data is given by the sum of a signal and a background contribution:
\begin{equation}
    f(x;\Gamma,\overbrace{B,S,\vec{s},\vec{b}}^{\vec{\nu}})=B\cdot f_b(x;\vec{b})+S\cdot f_s(x;\vec{s}),
    \label{model}
\end{equation}
where $B$ is the number of predicted background counts, $S$ is the number of signal counts, $f_b(x;\vec{b})$ and $f_s(x;\vec{s})$ are the predicted shapes of the background and signal, depending on some nuisance parameters $\vec{b}$ and $\vec{s}$, respectively. The $0\nu\beta\beta$ signal is a monochromatic peak, broadened only by the detector's response. We assume that the signal shape is given by a Gaussian function centered at Q$_{\beta\beta}$, with a standard deviation given by the detector energy resolution. The background can be very well approximated by an analytic function locally around the $0\nu\beta\beta$ peak. In this work we consider a constant background.  We tested the effect of a full background shape obtained from Monte Carlo (MC) simulations, shown in Fig. \ref{fig:cupid_bkg}. We parametrise the background shape with an exponential function plus four Gaussians for $^{214}$Bi peaks at 2978.9, 3000, 3053.9 and 3081.7 keV. The study performed for the case of the discovery and exclusion sensitivity in a Frequentist framework showed that the difference in the results between a constant background and a background shape from simulations is negligible (see Sections  \ref{section:freq_discovery} and \ref{section:freq_exclusion}).


\begin{figure}
    \centering
    \includegraphics[width=0.45\textwidth]{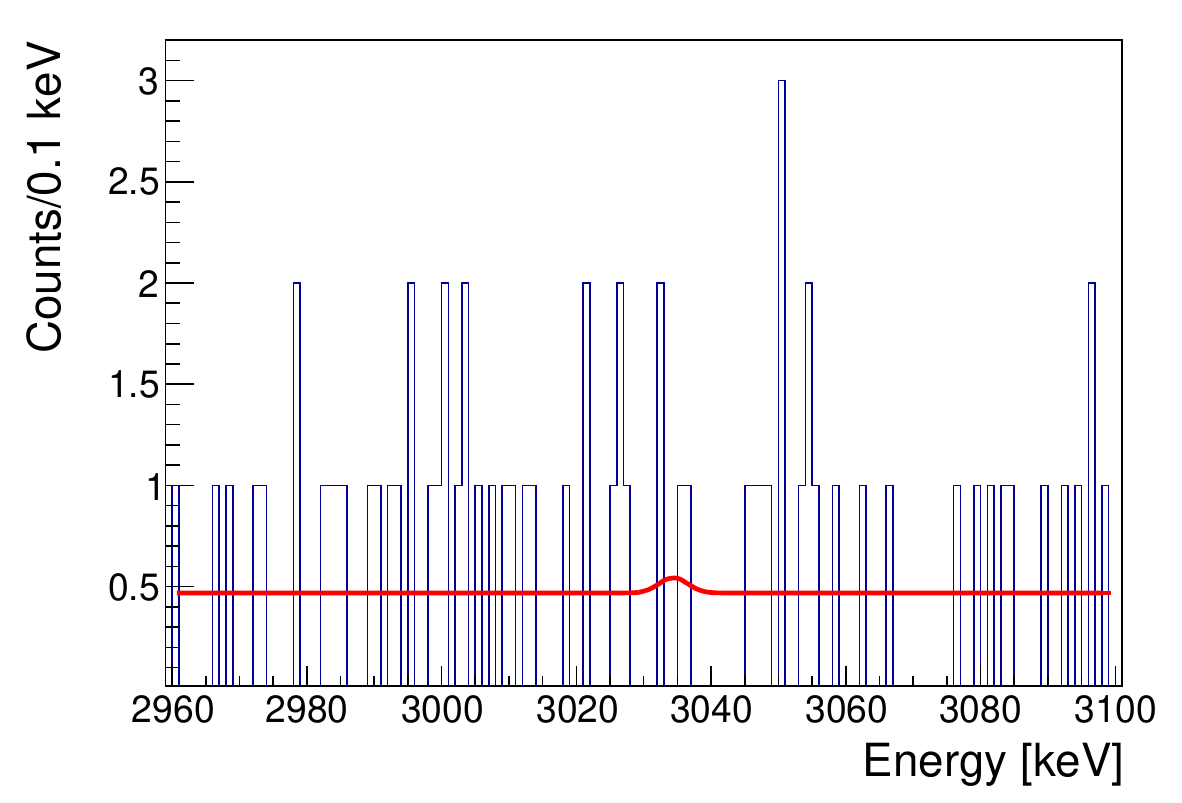}
    \includegraphics[width=0.45\textwidth]{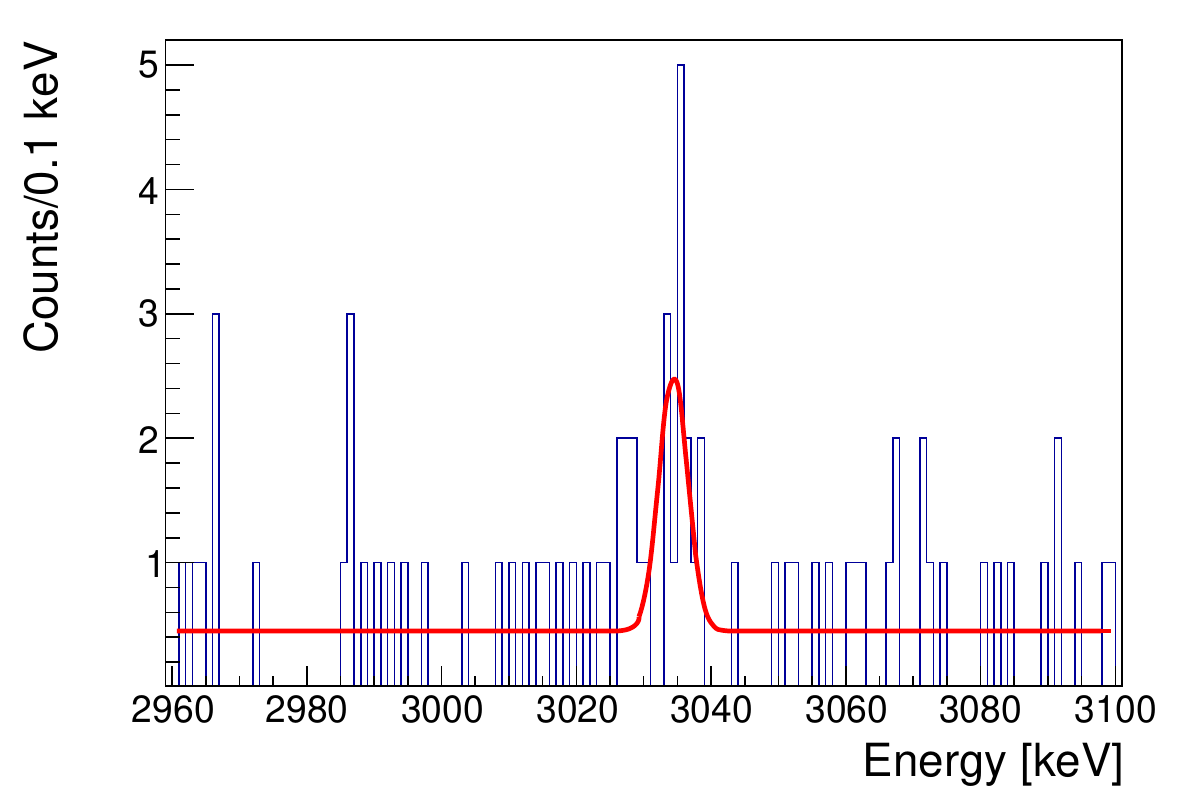}
    \caption{Example of two pseudo-experiments for the {\it baseline} background index of 1 $\times$ 10$^{-4}$ counts/keV/kg/yr. Top: with zero signal. Bottom: with signal rate  $\Gamma$ = $1\times 10^{-27}$ yr$^{-1}$. The best fit to Eq. \ref{model} is also shown.}
    \label{Eg}
\end{figure}

Throughout our analysis pseudo-experiments are required to generate distributions of test statistics, limits or discovery probabilities. These pseudo-experiments, or toy-MC experiments, are generated from Eq. \ref{model} for a given value of $B$ and $S$. Specifically, we sample from a Poisson distribution to obtain a number of background and signal events to generate in each experiment. The energies are then sampled from the probability distributions of the signal and background components. We generated pseudo-experiments with $B$ of $1.5 \times 10^{-4},1\times 10^{-4},6 \times 10^{-5}$ and $2 \times 10^{-5}$ counts/keV/kg/yr and with $\Gamma$ in the range $[0, 5 \times 10^{-27}$] yr$^{-1}$ in 100 steps. Examples of generated pseudo-experiments for the {\it baseline} $B = 1\times 10^{-4}$ counts/keV/kg/yr, and a signal rate of $\Gamma=0$ and $1\times 10^{-27}$ yr$^{-1}$ are shown in Fig. \ref{Eg}. 

\begin{figure}[h!]
    \centering
    \includegraphics[width=0.45\textwidth]{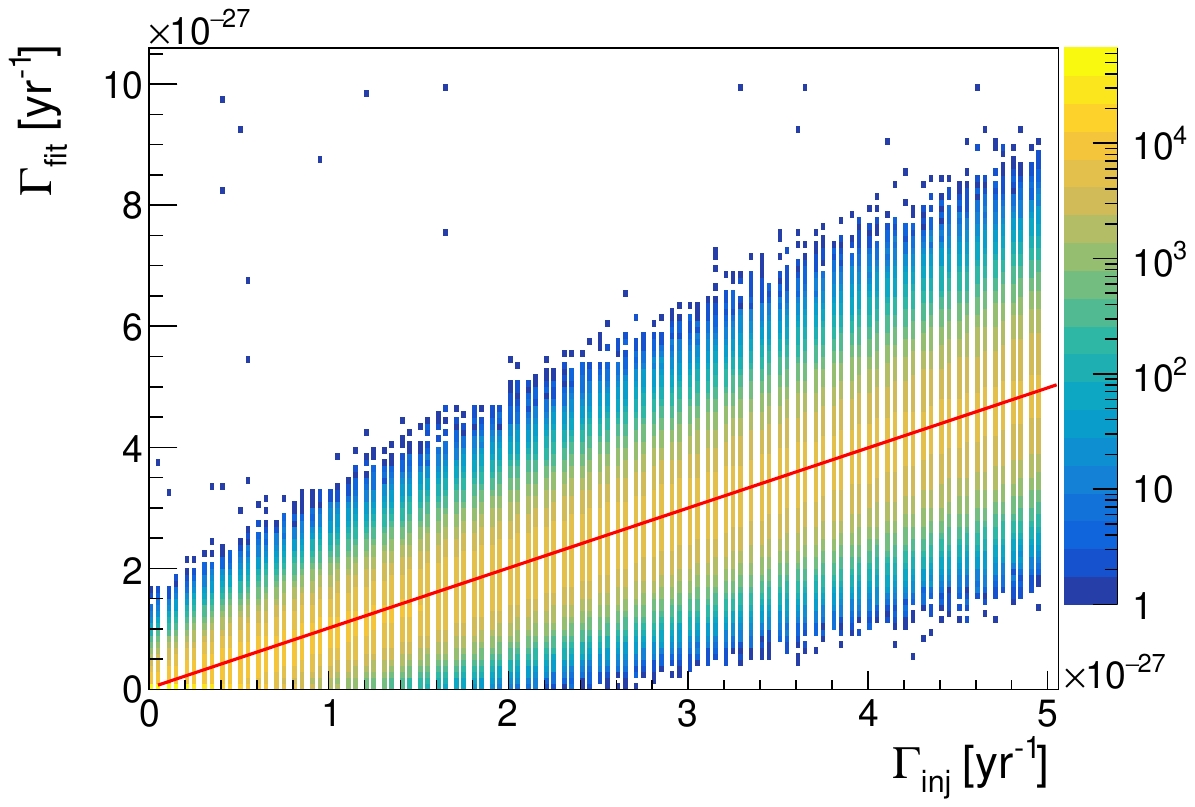}
    \caption{  
    Decay rates obtained from fitting pseudo-experiments with known decay rates, showing no significant bias in the reconstructed $\Gamma$ value. The background index is assumed to be  $1\times 10^{-4}$ counts/keV/kg/yr.}
    \label{bias}
\end{figure}

To study the bias in our procedure we generate and fit pseudo-experiments. 
   Figure  \ref{bias} shows the best fit versus the injected signal rate, $\Gamma$, for $B = 1.0\times 10^{-4}$ counts/keV/kg/yr.  
   We fit these distributions to a linear function to extract:
\begin{equation}
    \hat\Gamma_{\text{fit}} = 0.022(18) +0.992(6)\Gamma_\text{inj},
\end{equation}
which confirms there is no significant bias in our fitting procedure.


We also interpret the sensitivity in terms of $m_{\beta\beta}$. Under the hypothesis of light Majorana neutrino exchange, the half-life relates to the effective Majorana neutrino mass as:
\begin{equation}
    T_{1/2}^{-1} = g_A^4|M_{0\nu}^2| G_{0\nu}m_{\beta\beta}^2/m_e^2,
    \label{mbb}
\end{equation}
where $g_A$ is the axial-vector coupling constant, $G_{0\nu}$ is the phase factor, and $M_{0\nu}$ is the nuclear matrix element (NME) \cite{ParticleDataGroup:2024cfk}.

Thus, we use a set of NMEs and the phase factor to translate the sensitivity in $T_{1/2}^{-1}$ into a sensitivity in $m_{\beta\beta}$. We use the phase space factor computed in \cite{kot_psf}, $G_{0\nu}$ = 15.92 $\times$ 10$^{-15}$ yr$^{-1}$ and we set $g_A$ to the free nucleon value $g_A=1.27$. We consider the NMEs shown in Table~\ref{NMEs}, consistent with most recent reviews \cite{Agostini:2022zub,Gomez-Cadenas:2023vca}. We note that, as pointed out in \cite{shell_model}, the NME calculation within the shell model is particularly difficult for $^{100}$Mo and less NME calculations are reported for $^{100}$Mo compared to other candidate isotopes. In addition, remaining discrepancies between calculations and measurable observables like the $2\nu\beta\beta$ spectral shape \cite{CUPID-Mo:2023lru} and transition strength to excited states \cite{CUPID-Mo:2022cel} motivate further work to refine NME calculations. In the following, for the plots and tables, we will consider only the values from the references \cite{QRPA1, Deppisch:2020ztt, EDF2, shell_model}, for the sake of clarity.
Table~\ref{NMEs} gives the  {\it standard} NME, M$_L^{0\nu}$, calculated using long-range operators, which are typically considered by the community. It should be noted that most of the NME calculations of Gamow-Teller $\beta$ decays overestimate the experimentally measured ones \cite{Suhonen:2017krv}. This is often corrected by \emph{quenching} the value of the weak axial vector coupling constant, as an effective parameter \cite{Suhonen:2017krv, Suhonen:2013laa}. This disagreement can be due to the lack of nuclear correlations of the various many-body methods and/or the absence of two-body current operators in the models. These hypotheses tend to be confirmed by ab initio calculations, which naturally include these effects and were able to reproduce with good precision the experimental data \cite{Gysbers:2019uyb}. Nevertheless, it is still not certain how much it should impact the NME for the $0\nu\beta\beta$. Recent pn-QRPA calculations for $0\nu\beta\beta$ of $^{100}$Mo were done with two-body current, which lowered the NME by $\sim$(25--45)\% \cite{PhysRevC.107.044305}.

Some years ago studies based on the effective field theory demonstrated the need to introduce a leading-order short-range operator which gives rise to a short-range NME M$_S^{0\nu}$~\cite{PhysRevLett.120.202001,PhysRevC.97.065501,PhysRevC.100.055504}.
They are giving a sizable contribution to the total amplitude of the transition \cite{Gomez-Cadenas:2023vca, PhysRevC.107.044305, Castillo:2024jfj, Kauppinen:2025odt}, and in principle, as stressed in \cite{Kauppinen:2025odt}, their incorporation is relevant for experimental interpretations in $m_{\beta\beta}$. More generally, next-to-next leading order (N$^2$LO) NMEs should be considered, as pointed out in \cite{Castillo:2024jfj}, since their impact is small but non-negligible. So far, for $^{100}$Mo, the short-range NME has been calculated by the pn-QRPA \cite{PhysRevC.107.044305, Castillo:2024jfj} and IBM-2 models \cite{Kauppinen:2025odt}, which all agree on a substantial increase of the total NME. The extension to the N$^2$LO NMEs was performed in \cite{Castillo:2024jfj} for $^{100}$Mo, leading to an overall upper value of M$^{0\nu}$ = 10.91 when adding the uncertainties on each NME linearly. 
Nevertheless, the community is still considering only M$_L^{0\nu}$, and we decided to stick to this choice when quoting the default $m_{\beta\beta}$ ranges, for a fair comparison.


\begin{table}[h!]
    \centering
    \caption{Nuclear matrix elements used in the analysis of the present work. Only the references \cite{QRPA1, Deppisch:2020ztt, EDF2, shell_model} are used in the figures and tables as an example.}
    \begin{tabular}{ccc}
  Model & NME & Reference  \\ \hline \hline
    pn-QRPA & 3.90 & \cite{QRPA1} \\
    pn-QRPA & 5.868 & \cite{Simkovic:2018hiq} \\
    IBM-2 & 5.077 & \cite{Deppisch:2020ztt} \\
    EDF & 6.588 & \cite{EDF2} \\
    EDF & 5.08 & \cite{Rodriguez:2010mn} \\
    EDF & 6.48 & \cite{Song:2017ktj} \\
    Shell & 2.24 & \cite{shell_model}
    \end{tabular}
    \label{NMEs}
\end{table}

\section{Frequentist analysis}
\label{sec:freq}

In Frequentist statistics, the construction of confidence intervals and the estimation of the significance of a possible signal are based on hypothesis tests \cite{ParticleDataGroup:2024cfk,Cowan_freq,rec_DM}. We consider two hypotheses: a null hypothesis $H_0$, stating that some parameter is equal to a certain value and an alternative $H_1$ that this parameter is equal to any other value.

The optimal test statistic in a test of $H_0$ versus $H_1$ is given by the ratio of two likelihoods \cite{Feldman_cousins}:
\begin{equation}
    -2\ln{\frac{p(x|H_0)}{p(x|H_1)}}. 
\end{equation}
We evaluate $p(x|H)$ with the observed data $x$ and regard it as a function of the parameter(s) $\theta$: 
\begin{equation}
    t(\theta) =-2\ln{\frac{p(x|H_0)}{p(x|H_1)}}=-2\ln{\left(\frac{\mathcal{L}(\theta)}{\mathcal{L}(\hat{\theta})}\right)},
\end{equation}
where $\hat{\theta}$ is the value of the parameter $\theta$ maximizing the likelihood. An increasing value of $t(\theta)$ represents a decrease in the compatibility of the model with the null hypothesis.

For the case of a model with multiple parameters, of which only one is of interest, we can replace the likelihood function with the profile likelihood by maximizing the likelihood over the set of all nuisance parameters. In our case, $\hat{\vec{\nu}}$ are the set of nuisance parameters, related to the background level and the parameters of the signal and background shape, which maximize the likelihood for a given decay rate $\Gamma$:

\begin{equation}
    t_P(\Gamma) = -2\ln{\left(\frac{\mathcal{L}\big(\Gamma,\hat{\vec{\nu}}\big)}{\mathcal{L}(\hat{\Gamma},\hat{\vec{\nu}})}\right)}.
\end{equation}

\noindent By definition, $t_P(\Gamma)$ has a minimum at $\Gamma=\hat{\Gamma}$ with $t_P(\hat{\Gamma})=0$.

One complication which can arise is that, physically, the decay rate must be positive. In our work, we consider the rate $\hat{\Gamma}$ as an effective estimator allowed to be positive or negative. We then define our test statistic as:
\begin{equation}
\label{test_stat}
    t_P(\Gamma) = \begin{cases} -2\ln{\left(\frac{\mathcal{L}\left(\Gamma,\hat{\vec{\nu}}\right)}{\mathcal{L}\left(\hat{\Gamma},\hat{\vec{\nu}}\right)}\right)} & \hat{\Gamma}>0, \\
    \\
    -2\ln{\left(\frac{\mathcal{L}\left(\Gamma,\hat{\vec{\nu}}\right)}{\mathcal{L}\left(0,\hat{\vec{\nu}}(0)\right)}\right)} & \hat{\Gamma}<0.
    \end{cases}
\end{equation}
In this way, the test statistic for negative values of $\hat{\Gamma}$ is equal to that of $\hat{\Gamma}=0$, and the physical constraint on $\Gamma$ is satisfied.

\subsection{Discovery sensitivity}
\label{section:freq_discovery}

\begin{figure}[h!]
    \centering
      \includegraphics[width=0.45\textwidth]{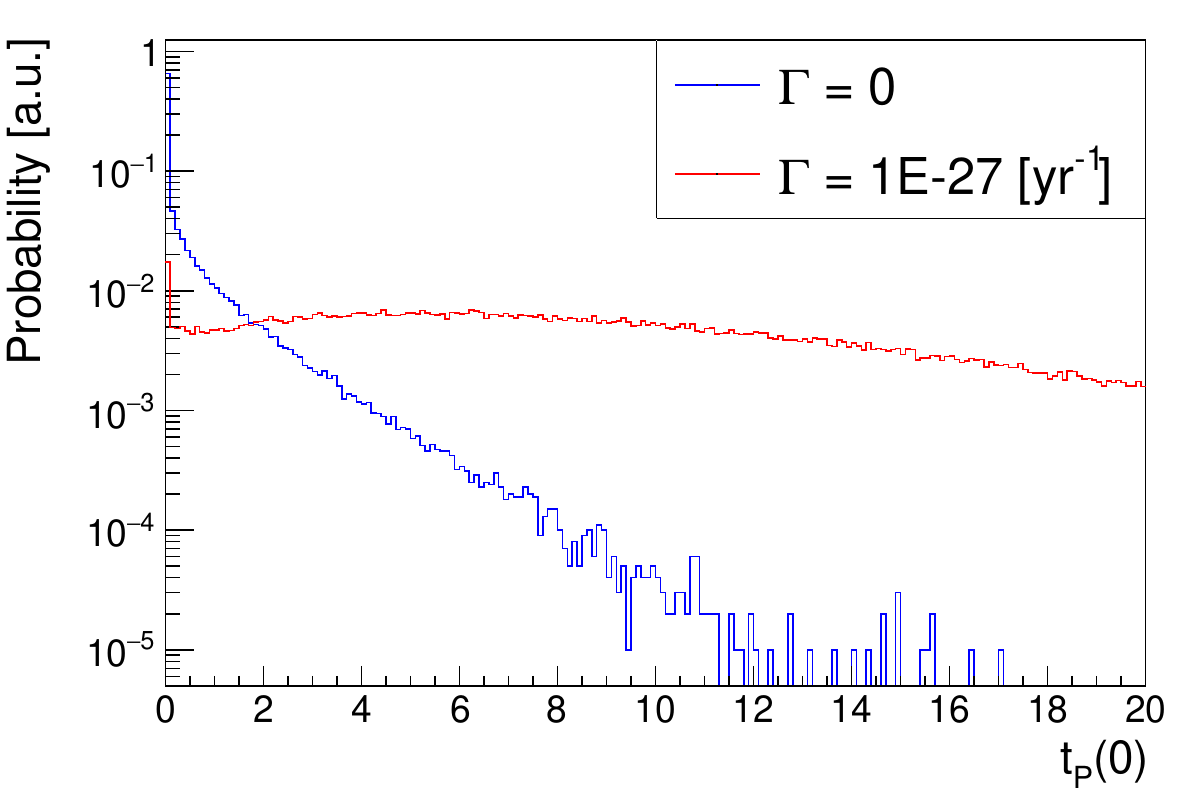}
    \caption{Probability distribution of the test statistic, $t_P(0)$ for zero decay rate in blue and $\Gamma=1 \times 10^{-27}$ yr$^{-1}$ in red, for $B$ = $10^{-4}$ counts/keV/kg/yr.}
    \label{tp0_distribution}
\end{figure}

Searching for a discovery or a claim of evidence is equivalent to a hypothesis test where the null hypothesis states the decay rate is equal to zero (background-only hypothesis) and the alternative that it is greater than zero (signal-plus-background). We consider the test statistic:
\begin{equation}
    t_P(0) =-2\ln{\left(\frac{\mathcal{L}(0)}{\mathcal{L}(\hat{\Gamma})}\right)},
    \label{eq:tp0}
\end{equation}
where $t_P(0)$ is small if the data agrees with the background-only hypothesis.
A claim of evidence ({\it discovery}) can be made if the value of the test statistic is greater than some cutoff. These cutoffs are sometimes estimated using asymptotic approximations \cite{Wilk}. However, this can lead to dramatically incorrect results if certain assumptions are not fulfilled. In this work, we instead estimate the distributions of the test statistic numerically using toy-MC experiments. In this way we ensure the correct coverage by construction. Figure \ref{tp0_distribution} shows an example of the probability distribution of the test statistic for $B$ = $10^{-4}$ counts/keV/kg/yr and zero decay rate and $\Gamma=1 \times 10^{-27}$ yr$^{-1}$.

\begin{figure}[h!]
    \centering
    \includegraphics[width=0.45\textwidth]{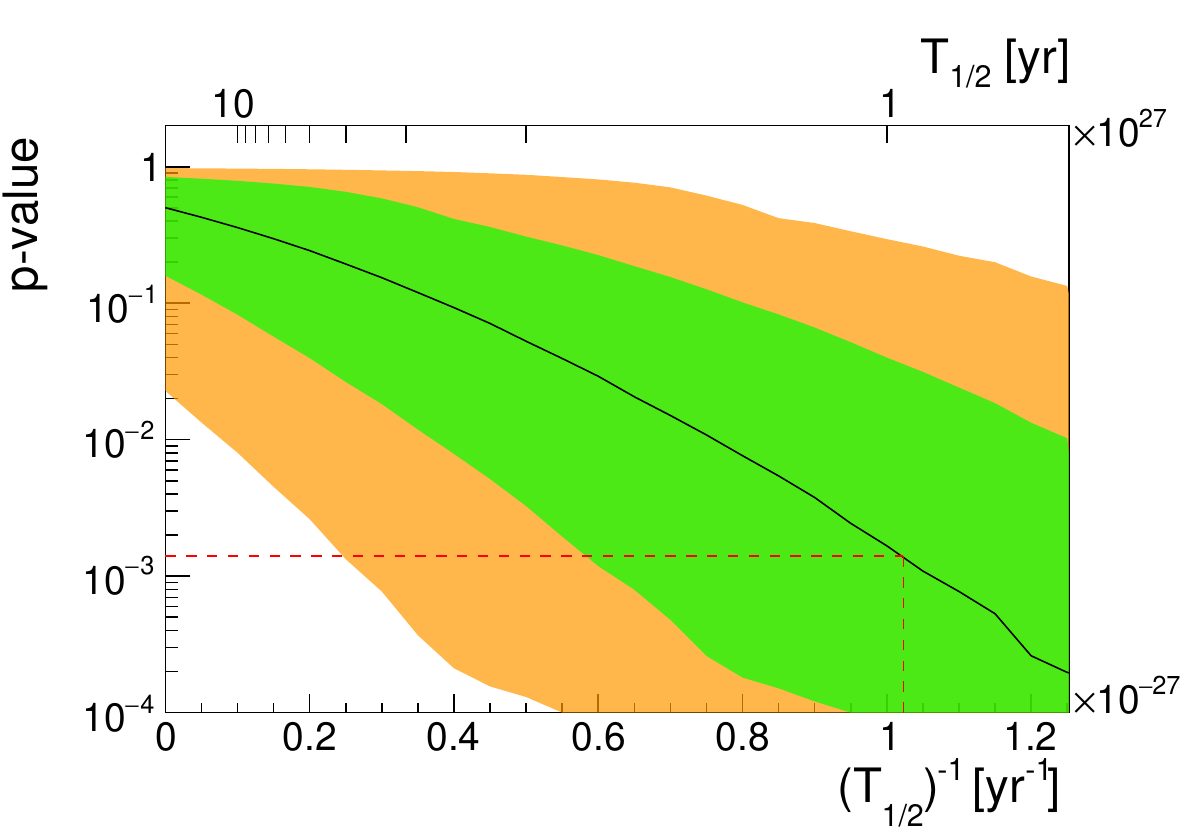}
    \caption{Background-only p-value, $p_b$, against the injected signal rate. We show the median with a black curve and 1, 2$\sigma$ bands in green and orange. The red dashed line highlights the rate value for which a 3$\sigma$ discovery is expected, corresponding to a p-value $p_b$ = 0.14\%.}
    \label{brazil_1}
\end{figure}

We compute the background-only p-value ($p_b$) as a function of the signal strength as:
\begin{equation}
\label{pb}
    p_b = \int_{t_P^{*}}^{\infty} f(t_P(0)|\Gamma=0) dt.
\end{equation}
Here $t_P^{*}$ is the value of the test statistic observed in each toy-MC experiment, $f(t_P(0)|\Gamma=0)$ is the probability distribution of the $t_P(0)$ test statistic generated with zero decay rate  shown in blue in Fig. \ref{tp0_distribution}. 
To compute numerically $f(t_P(0)|\Gamma=0)$, we generate an {\it array} of toy-MC experiments with zero signal.  We then generate separate arrays for different $\Gamma$'s and for each we extract $t_P(0)$, as defined in Eq. \ref{eq:tp0}. When $t_P(0)$ goes to higher values, it shows the incompatibility of the given pseudo-experiment with the background-only hypothesis. Finally, for each toy-MC experiment in the array of injected decay rates, we compute $p_b$ and we can build its probability distribution.  Figure \ref{brazil_1} shows the median of the background-only  $p_b$  distribution as a function of the decay rate  for $B = 10^{-4}$ counts/keV/kg/yr.
We estimate a median 3$\sigma$ discovery sensitivity by extracting the decay rate for $p_b = 0.0014$, and obtain $\hat{T}_{1/2}$ = 1.0 $^{+0.7}_{-0.3}$ $\times \ 10^{27} \ \mathrm{yr}$, where the uncertainty corresponds to the $\pm 34.1 \%$ interval around the median. 
The results for the three considered energy resolutions and the four background indices are given in Table~\ref{tab:discovery}. We point out that for the lowest $B = 0.2 \times 10^{-4}$ counts/keV/kg/yr, the low number of events leads to a discretisation in the probability distribution, thus the median sensitivity extracted can be quite different from the most probable value, as shown in the Appendix.
We translate the results on the half-life discovery sensitivity to  $m_{\beta\beta}$ values using Eq. \ref{mbb} and the NME's  in Table  \ref{NMEs}. The lowest values are obtained with the EDF model \cite{EDF2} and the upper values with the shell model \cite{shell_model}. For the CUPID \emph{baseline} scenario, this gives a range of $m_{\beta\beta}$ = 12.2--35.9 meV.

\begin{table}[h!]
    \centering
   \caption{CUPID  median discovery sensitivity ($3\sigma$) in half-life and effective Majorana neutrino mass, for 10 years livetime for various background indices and energy resolutions at \Qbb.}
    \begin{tabular}{c|c|c|c}
    Background index & Energy resolution &  $\hat{T}_{1/2}$ &   $\hat{m}_{\beta\beta}$  \\\
    [counts/keV/kg/yr] & FWHM [keV] & $10^{27}$ \ [$\mathrm{yr}]$ &  [$\mathrm{meV}$] \\ \hline 
    1.5 $\times 10^{-4}$ & 5   & 0.8 & 13--39   \\[3pt] \hline 
                         & 5   & 1.0  &  12--36  \\[3pt] 
    1.0 $\times 10^{-4}$ & 7.5 & 0.8  &  13--39  \\[3pt] 
                         & 10  & 0.7  &  14--42   \\[3pt] \hline 
    0.6 $\times 10^{-4}$ & 5   & 1.2 &  11--32  \\[3pt] \hline 
    0.2 $\times 10^{-4}$ & 5   & 1.7 &  9.3--27   \\ 
    \end{tabular}
    \label{tab:discovery}
\end{table}

\begin{figure*}[h!]
    \centering
    \includegraphics[width=0.45\textwidth]{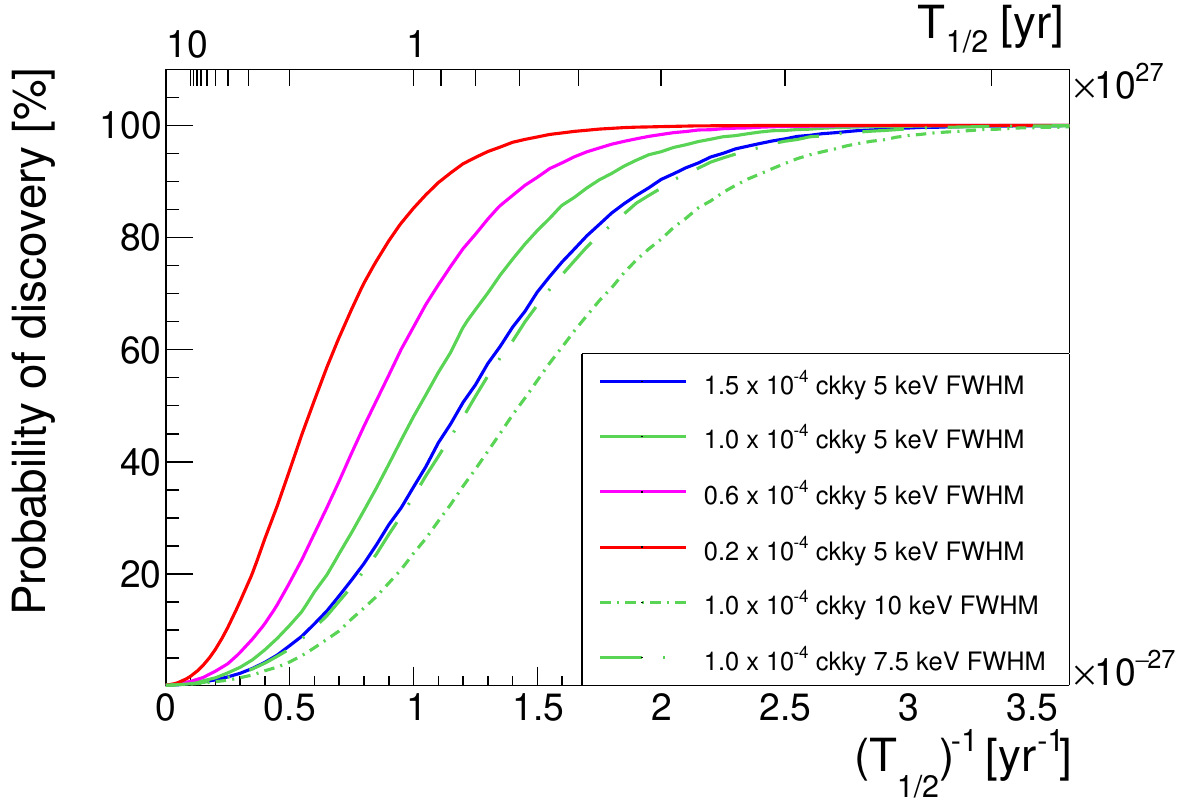}
      \includegraphics[width=0.45\textwidth]{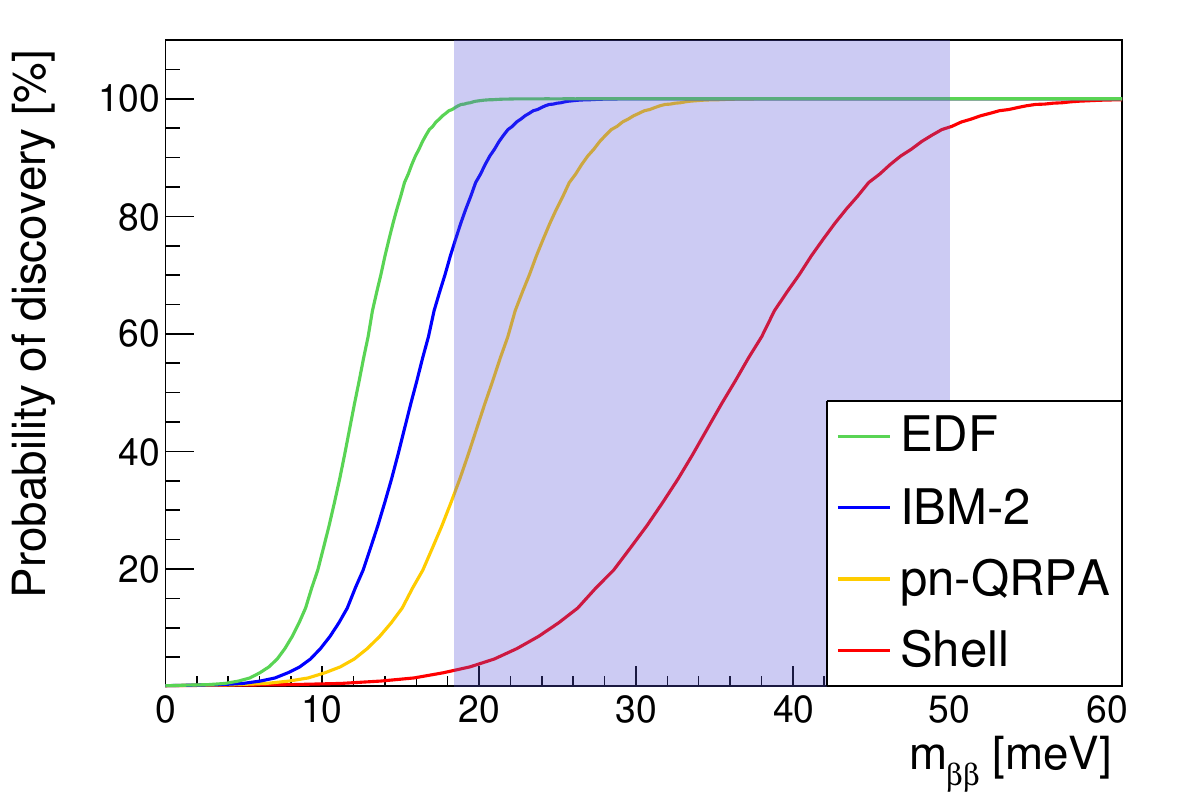}
    \caption{Left: Discovery probability for CUPID as a function of $T_{1/2}^{-1}$ for four considered background indices. Right: Discovery probability for the \emph{baseline} design ($B$ =  1 $\times 10^{-4}$ counts/keV/kg/yr, energy resolution of 5 keV FWHM) as a function of $m_{\beta\beta}$ for different nuclear models \cite{QRPA1, Deppisch:2020ztt, EDF2, shell_model}. The blue area corresponds to the inverted ordering region. A discovery is defined as the value at which there is 50\% probability of discovery.}
    \label{prob_discover}
\end{figure*}

We also compute the discovery probability as a function of the decay rate, shown in Fig. \ref{prob_discover}, left. We first produce the  probability distribution function of the discovery sensitivity, i.e. the decay rate for which $p_b = 0.0014$, in each toy-MC experiment. Then, the discovery probability is  computed by integrating this probability function. Figure \ref{prob_discover}, right, shows the corresponding discover probability for $B$ =  1 $\times 10^{-4}$ counts/keV/kg/yr and energy resolution of 5 keV (FWHM) as a function of $m_{\beta\beta}$, for the various nuclear models. For the most favourable nuclear models, a discovery, corresponding to the 50\% probability, is almost certain even for the smallest $m_{\beta\beta}$ values in the IO regime, shown as the blue area. The discovery probability for the effective Majorana neutrino mass at the bottom of the IO region, given by $m_{\beta\beta}=18.4$ meV, is shown for each model in Table \ref{tab:disc}.

\begin{table}[h!]
    \centering
   \caption{Discovery probability for CUPID \emph{baseline} design, given $m_{\beta\beta}$ at the bottom of the IO regime (18.4 meV) for various background indices in counts/keV/kg/yr and energy resolutions (FWHM), using the NMEs from \cite{QRPA1, Deppisch:2020ztt, EDF2, shell_model}.}
    \begin{tabular}{c|c|c|c|c|c}
    $B$ & $\Delta E$ & \multicolumn{4}{c}{\multirow{2}{*}{Probability [\%]}}\\ \
    [ckky] &  [keV] & \multicolumn{4}{c}{} \\ \hline
    & & IBM-2 & pn-QRPA  & EDF & Shell \\ \hline 
    1.5  $\times 10^{-4}$ & 5 & 62.9 & 23.0 & 96.1 & 1.9 \\[3pt] \hline
     & 5 & 75.2 & 32.7 & 98.4 & 2.7 \\[3pt]
    1.0 $\times 10^{-4}$ & 7.5 & 60.6 & 21.5 & 95.3 & 1.8 \\[3pt]
     & 10 & 47.3 & 14.6 & 90.0 & 1.2 \\[3pt] \hline
    0.6 $\times 10^{-4}$ & 5 & 86.9 & 47.9  & 99.5 & 4.8 \\[3pt] \hline
    0.2 $\times 10^{-4}$ & 5 & 96.7 & 73.2  & 99.96 & 12.3
    \end{tabular}
    \label{tab:disc}
\end{table}

We investigated the effect of using a background shape derived from the MC simulations rather than a constant background, as outlined in Section \ref{sec:stat}. We focused on $B$ = 1.0~$\times~10^{-4}$ counts/keV/kg/yr and considered the extreme worst case scenario of a 10 keV energy resolution FWHM at \Qbb. We obtained $\hat{T}_{1/2}$  = 0.66 $\times$ $10^{27}$ yr, which is within the uncertainties of the result with a flat background, 0.70$^{+0.48}_{-0.23} \ \times$ $10^{27}$ yr.

\subsection{Exclusion sensitivity}
\label{section:freq_exclusion}
The exclusion sensitivity is computed similarly, but in this case, we perform a hypothesis test comparing a null hypothesis that the decay rate is equal to the signal $S$ in our model (see Eq. \ref{model}), $\Gamma=S$, to an alternative that is $\Gamma\neq S$. The exclusion limit is obtained when the null hypothesis is incompatible with the given toy-MC experiment. We test each $\Gamma$ as the null hypothesis: for each decay rate, we look at how compatible are the other decay rate hypotheses with this one. A decay rate can be excluded if it is sufficiently unlikely for toy-MC experiments generated with this decay rate to produce the data.

We compute the profile likelihood ratio, $t_P(\Gamma)$, for each toy-MC experiment. We also compute the $p(t_P(\Gamma))$ distribution for toy-MC-generated with this value of $\Gamma$. From this, we obtain:
\begin{equation}
    p_\mu(\Gamma)= \int_{t_P}^{\infty} p(t_P(\Gamma)|\Gamma) dt,
\end{equation}
where $t_P$ is the value of $t_P(\Gamma)$ obtained in a given toy-MC experiment. From this, we can compute the 90\% confidence level (CL) as the interval of $\Gamma$ values with $p_\mu(\Gamma)>$ 0.1. In our analysis of exclusion sensitivity, we only present one-sided intervals. This choice leads to some moderate over-coverage of our intervals in some cases.

\begin{figure}[h!]
    \centering
    \includegraphics[width=0.45\textwidth]{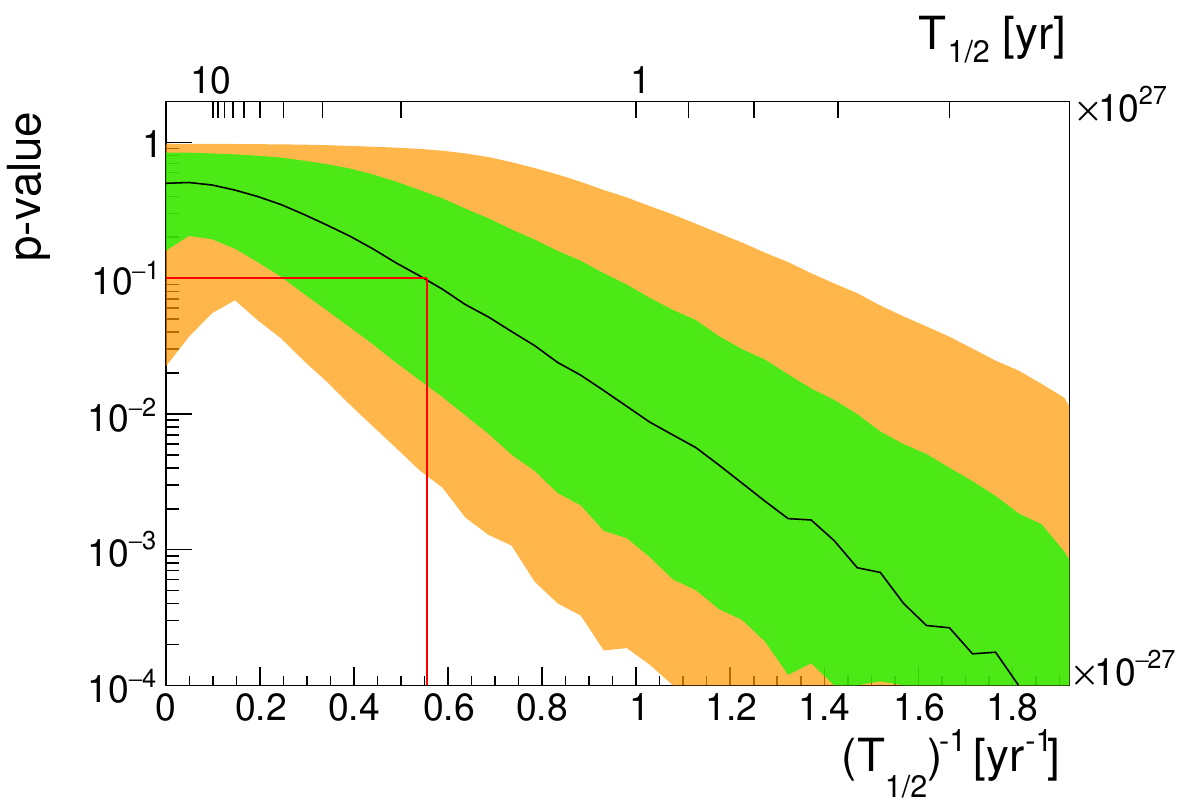}
    \caption{Distribution of $p_{\mu}({\Gamma})$, obtained by integrating the profile likelihood ratio for a given decay rate ${\Gamma}$, as a function of the injected signal rate, for the CUPID \emph{baseline} design ($B$ = 1.0 $\times$ 10$^{-4}$ \ $\mathrm{counts/keV/kg/yr}$ and energy resolution of 5 keV). We show the median of the distribution (black line) and 1, 2$\sigma$ confidence bands. The red line represents the rate value for which $p_\mu$ = 0.1, corresponding to the 90\% CL exclusion.}
    \label{pmu}
\end{figure}

Figure \ref{pmu} shows the distribution of $p_\mu$ for the CUPID \emph{baseline} design. 
Based on the threshold of $p<10\%$ for 90\% CL exclusion, we extract the median exclusion sensitivity of $\hat{T}_{1/2}$ $>$ 1.8 $^{+2.1}_{-0.8}$ $\times \ 10^{27} \ \mathrm{yr}$, where the uncertainty is given by the range of $\pm 34.1 \%$ around the median.

In addition, for each toy-MC, we find a curve of $p_\mu({\Gamma})$ against $(T_{1/2})^{-1}$. The interception of this line with 10\% gives the limit which would be obtained for this pseudo-experiment. The corresponding  distribution of limits is shown in the Appendix. We extract a median Frequentist 90\% CL exclusion sensitivity for the various background indices and energy resolutions given in Table \ref{tab:exclusion}. 
Translating in terms of the effective Majorana mass, the exclusion sensitivity yields  $m_{\beta\beta}$ $<$ 9.0--26 meV for $B$ = 1.0~$\times~10^{-4} \ \mathrm{counts/keV/kg/yr}$ and 5 keV FWHM energy resolution.

We also investigated the effect of using a background shape based on the MC simulations instead of a constant background (Section \ref{sec:stat}). For the CUPID {\it baseline} scenario, 
we obtained $\hat{T}_{1/2}$  = 1.81 $\times$ $10^{27}$ yr, which is in line to the result obtained  with the flat background  (1.80 $\times$ $10^{27}$ yr).

\section{Bayesian analysis}
\label{sec:Bay}

Bayesian statistical methods start from a different interpretation of probability. In Bayesian statistics, probability represents the degree of belief in the true value of a parameter or a model \cite{bayes}. Starting from some prior degree of belief about a set of parameters $\pi(\vec{\theta)}$, Bayesian methods utilize some data $\mathcal{D}$ to update this degree of belief to obtain the posterior probability distribution $p(\vec{\theta}|\mathcal{D})$. This is achieved using the Bayes theorem:
\begin{equation}
    p(\vec{\theta}|\mathcal{D})=\frac{p(\mathcal{D}|\vec{\theta})\pi(\vec{\theta})}{p(\mathcal{D})}.
\end{equation}
Here $p(\mathcal{D})$ does not depend on $\vec{\theta}$ and can thus be considered as a normalization constant and $p(\mathcal{D}|\vec{\theta})$ is the likelihood function defined earlier in Eq. \ref{likelihood}. Bayesian inferences are based on this probability distribution.

\subsection{Discovery sensitivity}

To compute a Bayesian discovery sensitivity, we need to compare two models, one where the decay rate $\Gamma=0$ and another with $\Gamma\neq 0$.
Bayesian model comparison is based on comparing the posterior probabilities of multiple hypotheses \cite{model_1,model_2}. A Bayesian discovery is claimed when the null hypothesis has a sufficiently low posterior probability. For two models $H_0$ and $H_1$:
\begin{equation}
    p(H_i|\mathcal{D}) = \frac{p(\mathcal{D}|H_i)p(H_i)}{p(\mathcal{D}|H_0)p(H_0)+p(\mathcal{D}|H_1)p(H_1)}.
\end{equation}
We can reject the null hypothesis $H_0$ if $p(H_0|\mathcal{D})$ is less than some cutoff. This is equivalent to requiring a Bayes factor, defined as:
\begin{equation}
B(H_0,H_1) = \frac{p(\mathcal{D}|H_1)}{p(\mathcal{D}|H_0)}\frac{p(H_1)}{p(H_0)},
\end{equation}
to be greater than some cutoff. It is typical to choose the two models $H_0$ and $H_1$ to be equally likely {\it a priori}. Therefore, the problem of Bayesian model comparison reduces to the integration of the posterior distribution over the full parameter space for several models. However, in this way, the evidence can depend strongly on the choice of priors on the model parameters. The natural solution  is to use instead an informative prior on the signal $S$. This study of the dependence of  the discovery sensitivity on the priors goes beyond the scope of this work.  The dependence on the Bayesian priors along with ways of reducing it is studied separately for both NO and IO in a global analysis in  \cite{discovery}. 

\subsection{Exclusion sensitivity}

To compute the experimental exclusion sensitivity rather than model comparison, we compute credible intervals (c.i.). 
We obtain the marginalized posterior distribution by integrating all the nuisance parameters $\vec{\nu}$ over their parameter space $\Omega$:
\begin{equation}
    p(\Gamma|\mathcal{D}) =\int_{\Omega}p(\Gamma,\vec{\nu}|\mathcal{D})d\vec{\nu}.
\end{equation}

From this distribution, we can obtain Bayesian credible intervals, which contain the true value of the parameter with a certain probability. In our analysis, we quote $90$\% c.i. upper limits on the $0\nu\beta\beta$ decay rate. 

We use Markov Chain Monte Carlo (MCMC), implemented with the Bayesian Analysis Toolkit (BAT) \cite{BAT}, to sample the full posterior distribution of the parameters. 

Our analysis considers a flat prior on the decay rate, or equivalently  $m_{\beta\beta}^2$. This choice is standard in the $0\nu\beta\beta$ community \cite{GERDA:2020xhi,CUORE_nature} and is seen as a conservative choice. Other possible priors include one flat on $m_{\beta\beta}$ or log$(m_{\beta\beta}$). These would typically give stronger limits.

\begin{figure*}[h!]
    \centering
    \includegraphics[width=0.45\textwidth]{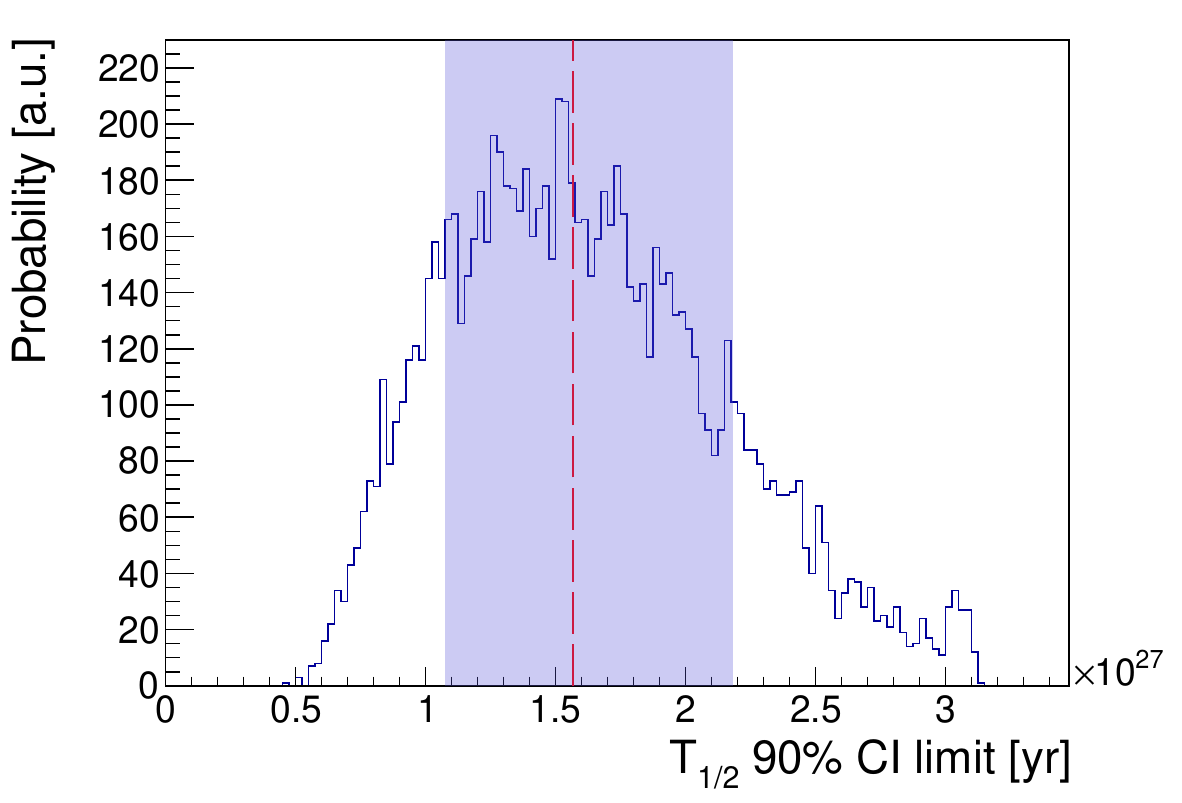}
\includegraphics[width=0.45\textwidth]{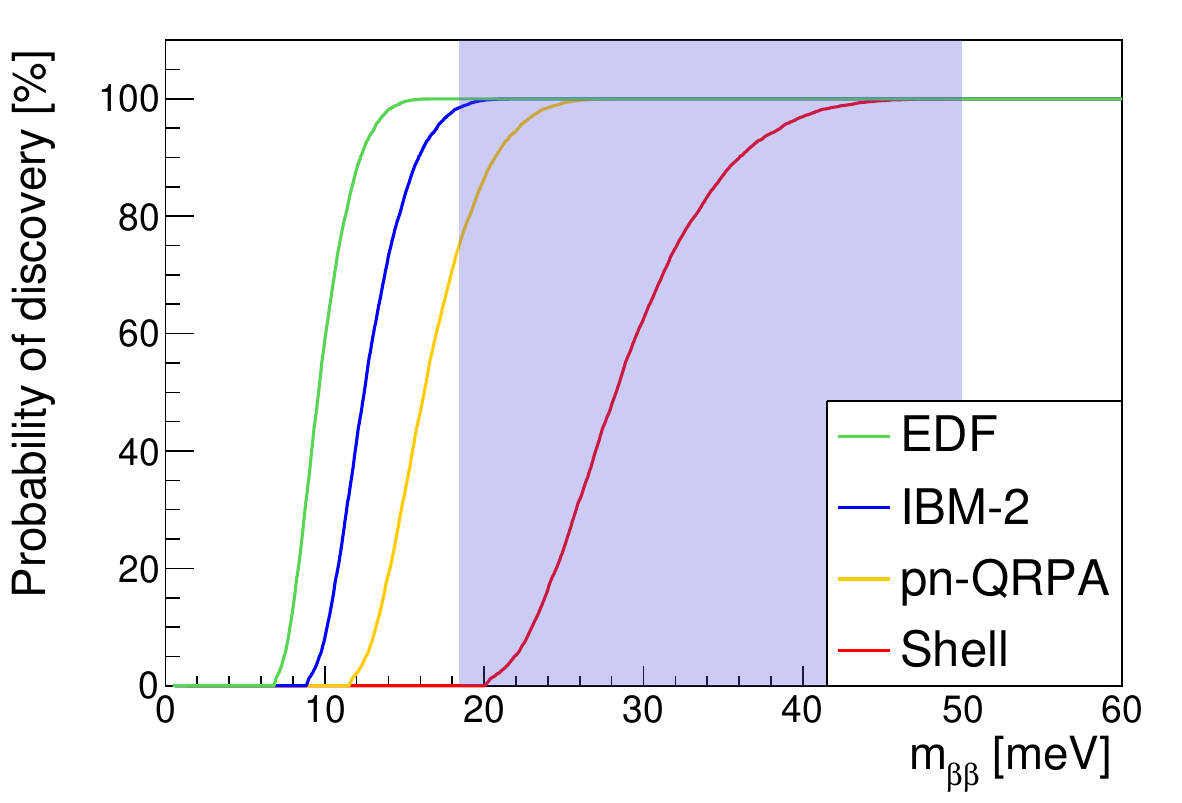}

    \caption{Probability of 90\% exclusion limits for CUPID obtained with a Bayesian analysis, under the assumption of a background of $1\times 10^{-4}$~counts/keV/kg/yr and 5 keV FWHM energy resolution. Left: Distribution of half-life exclusion limits, the median sensitivity and interval containing 68\% of the pseudo-experiments are shown. 
   Right: Probability for CUPID to exclude $m_{\beta\beta}$ values at 90\% c.i. The probability of 50\% corresponds to the median of the distributions on the left translated in $m_{\beta\beta}$. The shaded band corresponds to the inverted ordering region.
    }
    \label{dist}
\end{figure*}

We  extract the marginalized posterior distribution on $T_{1/2}^{-1}$ and thus the 90\% c.i. upper limit on $T_{1/2}^{-1}$ for every pseudo-experiment. In Fig. \ref{dist}, left, we show the distribution of limits assuming a  background of 1 $\times \ 10^{-4}$ counts/keV/kg/yr. We highlight the median sensitivity and a band containing 68\% of the pseudo-experiments. 
The median 90\% c.i. exclusion sensitivities are given in Table \ref{tab:exclusion} 
for several background and energy resolution scenarios. 
We convert the distribution of excluded $T_{1/2}$ into a distribution on limits on $m_{\beta\beta}$ for each NME in Table \ref{NMEs}.  
The median 90\% c.i. exclusion sensitivities on $m_{\beta\beta}$ obtained from these distributions are also shown in Table \ref{tab:exclusion}.


\begin{table}[h!]
    \centering
   \caption{CUPID median exclusion sensitivity at 90\% confidence limit in half-life and effective Majorana neutrino mass, for 10 years livetime, computed in the Bayesian and Frequentist analyses for four different background indices in counts/keV/kg/yr and three considered energy resolution (FWHM) at \Qbb.}
    \begin{tabular}{c|c|c|c|c|c}
    $B$ & $\Delta E$ &  \multicolumn{2}{c}{$\hat{T}_{1/2}$} &  \multicolumn{2}{|c}{$\hat{m}_{\beta\beta}$}  \\\
    [ckky] & [keV] & \multicolumn{2}{c}{$10^{27}$ \ [$\mathrm{yr}]$} &  \multicolumn{2}{|c}{[$\mathrm{meV}$]} \\ \hline 
    & & Bay. & Freq. & Bay. & Freq. \\ \hline
    1.5 $\times 10^{-4}$ & 5   & 1.4 & 1.5 & 10--30  & 9.7--29   \\[3pt] \hline 
                         & 5   & 1.6 & 1.8 &  9.6--28  &  9.0--26 \\[3pt] 
    1.0 $\times 10^{-4}$ & 7.5 & 1.4 & 1.5 &  10--30 &  9.7--29 \\[3pt] 
                         & 10  & 1.2 & 1.3 & 11--32  & 10--31  \\[3pt] \hline 
    0.6 $\times 10^{-4}$ & 5   & 1.8 & 2.2 &   8.9--26 &  8.2--24 \\[3pt] \hline 
    0.2 $\times 10^{-4}$ & 5   & 2.3 & 3.1 &  7.9--23  &  6.9--20  \\ 
    \end{tabular}
       \label{tab:exclusion}
\end{table}

We also compute the probability to exclude a $m_{\beta\beta}$ value at 90\% c.i. as a function of $m_{\beta\beta}$,  shown in Fig.  \ref{dist}, right, for the  CUPID \emph{baseline} 
This is obtained by integrating the distributions of the  $m_{\beta\beta}$ exclusion limits. The probability to fully exclude the IO regime is $\sim$1 for all considered background indices for the best-case NMEs.

\section{Staged deployment}
CUPID considers deploying the detector in two phases enabling early science data. In a possible scenario the first phase of the experiment -- CUPID Stage-I --  would deploy 1/3 of the detectors arranged in 19 towers. While the additional towers are being built, we would be able to take data with the first phase of the experiment. After several years of data taking, we would incorporate the additional planned crystals. 

We calculated the discovery sensitivity in the Frequentist framework for CUPID Stage-I assuming 3 years of data taking. We adopted a first stage scenario consisting in the deployment of 1/3 of the detectors with  150 kg of the total crystal mass. 
As for the full CUPID, we studied the impact on the discovery sensitivity of the energy resolution and consider 5 keV, 7.5 keV and 10 keV FWHM at \Qbb. In CUPID Stage-I we expect a slightly higher background with respect to the full experiment due to the smaller number of detectors, which reduces the events in  coincidences and therefore the tagging of muons and of the contaminations in the crystals. Current studies of the background in CUPID Stage-I indicate 
$B \sim 1.2 \times 10^{-4}$ counts/keV/kg/yr, thus we consider a conservative scenario with 1.5 $\times$ 10$^{-4}$ counts/keV/kg/yr and another one with $B$ = 1.0 $\times$  10$^{-4}$ counts/keV/kg/yr. 
Table \ref{tab:discovery_resolution_PhaseI} gives the resulting discovery sensitivities for CUPID Stage-I.


\begin{table}[h!]
    \centering
   \caption{CUPID Stage-I median discovery sensitivity ($3\sigma$) in half-life and effective Majorana neutrino mass, for 3 years livetime and the crystal mass of 150 kg for two different background indices and three considered energy resolutions at \Qbb.}
    \begin{tabular}{c|c|c|c}
    Background index & Energy resolution &  $\hat{T}_{1/2}$ &   $\hat{m}_{\beta\beta}$  \\\   
    [counts/keV/kg/yr] & FWHM [keV] & $10^{27}$ \ [$\mathrm{yr}]$ &  [$\mathrm{meV}$] \\ \hline 
    
         & 5 & 0.21 & 26--77   \\[3pt]  
     1.0 $\times 10^{-4}$ & 7.5 & 0.18 & 29--85  \\[3pt] 
         & 10 & 0.15  &  31--91  \\[3pt] \hline 

         & 5 & 0.18 & 28--83   \\ 
     1.5 $\times 10^{-4}$ & 7.5 & 0.16 & 30--89  \\[3pt] 
         & 10 & 0.14  &  33--96  \\[3pt] 
    \end{tabular}
       \label{tab:discovery_resolution_PhaseI}
\end{table}

Assuming the {\it baseline} design parameters, with energy resolution of 5 keV and $B = 1 \times 10^{-4}$ counts/keV/kg/yr,  in CUPID Stage-I we get a median discovery sensitivity of 0.2~$\times~10^{27}$ yr, corresponding to a range of  $m_{\beta\beta}$ = 26 -- 77 meV. In this scenario, the staged deployment would add an exposure of 3 years $\times$ 150 kg, allowing to reach the final sensitivity with an additional 9 years of data taking with the full CUPID.

\begin{figure}[h!]
    \centering
    \includegraphics[width=0.45\textwidth]{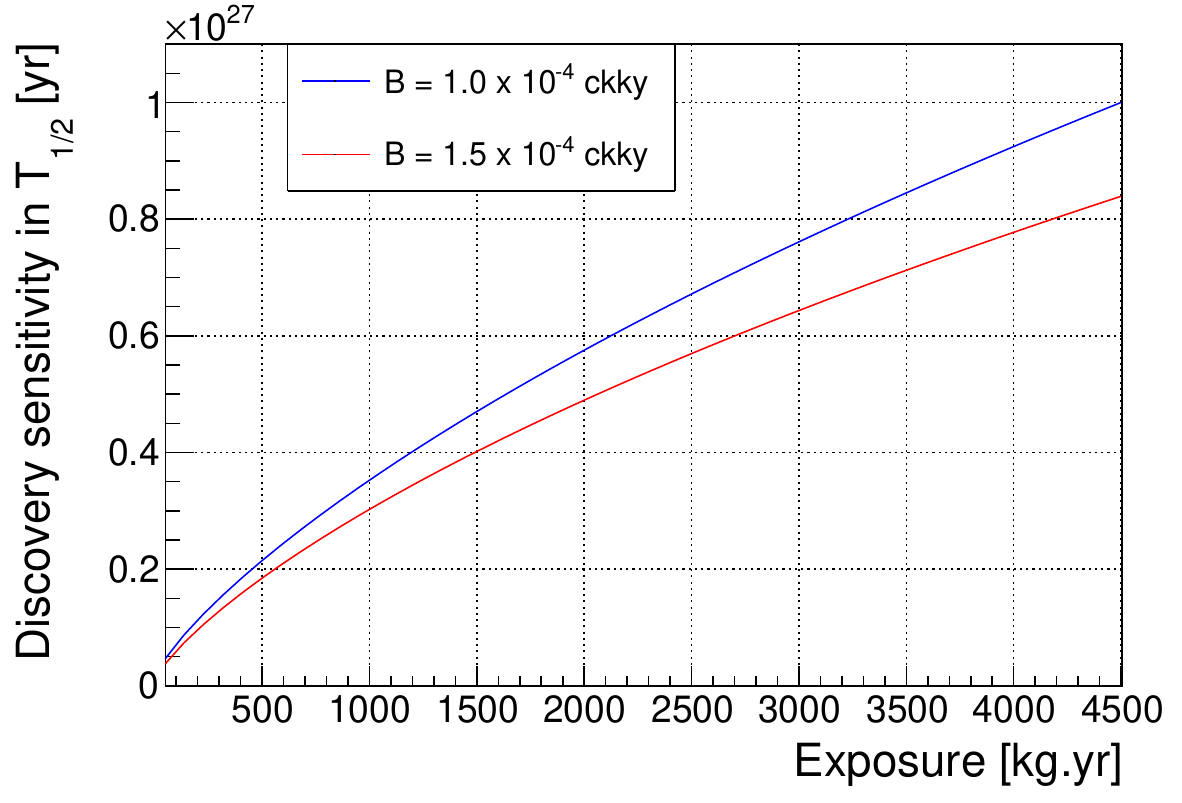}
    \caption{CUPID median discovery sensitivity ($3\sigma$) in half-life as a function of the exposure in crystal mass times livetime. The sensitivity for $B = 1 \times 10^{-4}$ counts/keV/kg/yr and $1.5 \times 10^{-4}$ counts/keV/kg/yr are shown in blue and red respectively. This is obtained using an energy resolution of 5 keV FWHM.}
    \label{fig:expo_disco}
\end{figure}

The evolution of the discovery sensitivity as a function of the exposure for $B = 1.0 \times 10^{-4}$ counts/keV/kg/yr and $1.5 \times 10^{-4}$ counts/keV/kg/yr are shown in Fig. \ref{fig:expo_disco}.

\section{Conclusion}

\begin{figure}[h!]
    \centering
    \includegraphics[width=0.45\textwidth]{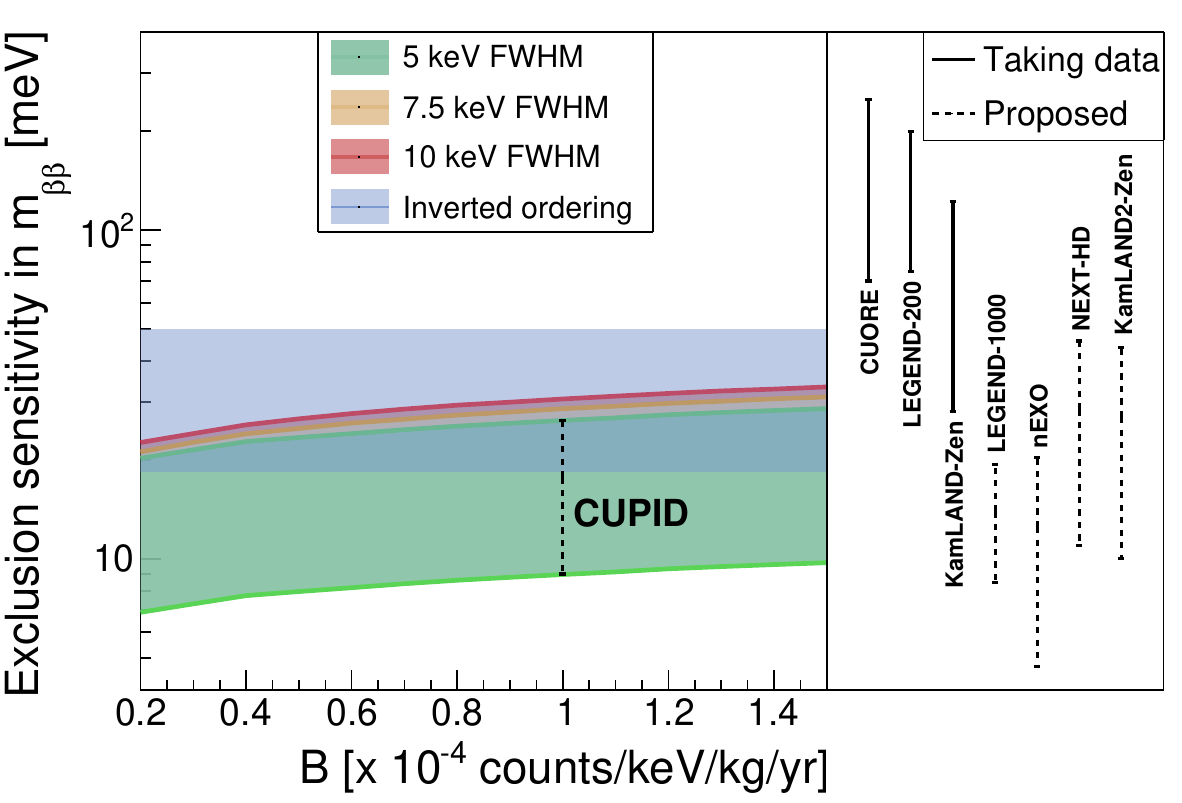}
    \caption{90 \% c.i. exclusion sensitivity in $m_{\beta\beta}$ as a function of the background index in CUPID. The bands in green, orange, and red show the range of $m_{\beta\beta}$ values, with NME's from Table \ref{NMEs}, for an energy resolution of 5 keV, 7.5 keV, and 10 keV, respectively. We highlight the CUPID \emph{baseline} scenario with 5 keV energy resolution and a background index of $1.0 \times 10^{-4}$ counts/keV/kg/yr. The blue band represents the inverted ordering region. The right panel shows the exclusion limits of current \cite{CUORE:2024ikf, 25tk-nctn, KamLAND-Zen:2024eml} (solid) and future experiments \cite{LEGEND:2021bnm, nEXO:2021ujk, NEXT:2020amj, KamLand2Zen} for a data taking of 10 years (dashed).}
    \label{fig:summary}
\end{figure}

In this work we computed the expected sensitivity of the next-generation experiment CUPID to search for neutrinoless double-beta decay of $^{100}$Mo. In the CUPID \emph{baseline} scenario, with crystal mass 450 kg, the background is  1.0~$\times~10^{-4}$ counts/keV/kg/yr and the energy resolution is 5 keV FWHM at the Q-value of the transition. Using an  unbinned likelihood analysis and a Frequentist and Bayesian frameworks, we obtain the discovery and the exclusion sensitivities to the half-life and the effective Majorana neutrino mass. 
In the \emph{baseline} scenario, with 10 live-years of data taking, we derive a $3\sigma$ discovery sensitivity in the Frequentist approach of $\hat{T}_{1/2}$ = 1.0 $^{+0.7}_{-0.3}$ $\times \ 10^{27} \ \mathrm{yr}$, corresponding to $\hat{m}_{\beta\beta}= \ 12$--$36 \ \mathrm{meV}$. The uncertainty given on the half-life in the Frequentist discovery sensitivity, as well as for the Frequentist and Bayesian exclusion sensitivities, is given by the $\pm 34.1 \%$ interval around the median, and is not taken into account when quoting the $\hat{m}_{\beta\beta}$ ranges. Without the value of the shell model NME \cite{shell_model}, the CUPID sensitivity  becomes $\hat{m}_{\beta\beta}= \ 12$--$21 \  \mathrm{meV}$.
On the other hand, accounting for the short-range M$_S^{0\nu}$ and the N$^2$LO NMEs \cite{Castillo:2024jfj} results in a smaller lower value of 7.4 meV.
The median exclusion sensitivity in the Frequentist analysis yields $\hat{T}_{1/2} = 1.8^{+2.1}_{-0.8} \times 10^{27} \ \mathrm{yr}$, corresponding to exclusion limits to the effective Majorana neutrino mass of  $\hat{m}_{\beta\beta}= \ 9.0$--$26 \ \mathrm{meV}$. The  Bayesian exclusion limit  at  90\% c.i. results in $\hat{T}_{1/2} = 1.6^{+0.6}_{-0.5} \times 10^{27} \ \mathrm{yr}$, corresponding to limits to the effective Majorana neutrino mass of  $\hat{m}_{\beta\beta}= \ 9.6$--$28 \ \mathrm{meV}$, covering the neutrino mass regime in the inverted ordering  scenario, as well as the normal ordering  regime with neutrino mass larger than 10~meV. 
We have also calculated the impact on the expected sensitivity in the case of background levels and energy resolutions differing from those of the \emph{baseline} scenario. The power of discovery of the CUPID experiment is highlighted in Fig. \ref{fig:summary}, which shows the discovery sensitivity on the effective Majorana neutrino mass as a function of the background index for various energy resolutions. The bands represent the range of $m_{\beta\beta}$ values using the NMEs from Table \ref{NMEs}. The variations on the discovery sensitivity in $m_{\beta\beta}$ are dominated by the spread in NME values, rather than by the differences induced by the various scenarios of background index and energy resolution. The right panel in Fig. \ref{fig:summary} shows the comparison with current and next-generation experiments, underlying the competitiveness of CUPID.

\section{Acknowledgments}
The CUPID Collaboration thanks the directors and staff of the Laboratori Nazionali del Gran Sasso and
the technical staff of our laboratories. This work was supported by the Istituto Nazionale di Fisica Nucleare (INFN); by the European Research Council (ERC) under the European Union Horizon 2020 program (H2020/2014-2020) with the ERC Advanced Grant no. 742345 (ERC-2016-ADG, project CROSS) and the Marie Sklodowska-Curie Grant Agreement No. 754496; by the Italian Ministry of University and Research (MIUR) through the grant Progetti di ricerca di Rilevante Interesse Nazionale (PRIN)  grant no. 2017FJZMCJ and grant no. 2020H5L338;  by the Agence Nationale de la Recherche (ANR) in France, through the  ANR-21-CE31-0014- CUPID-1; by the US National Science Foundation under Grant Nos. NSF-PHY-1401832, NSF-PHY-1614611, NSF-PHY-2412377 and NSF-PHY-1913374. This material is also based upon work supported by the US Department of Energy (DOE) Office of Science under Contract Nos. DE-AC02-05CH11231 and DE-AC02-06CH11357; and by the DOE Office of Science, Office of Nuclear Physics under Contract Nos. DE-FG02-08ER41551, DE-SC0011091, DE-SC0012654, DE-SC0019316, DE-SC0019368, and DE-SC0020423. This work was also supported by the Russian Science Foundation under grant No. 18-12-00003. 

\bibliographystyle{spphys}       
\bibliography{biblio}

@article{Fukugita:1986hr,
    author = "Fukugita, M. and Yanagida, T.",
    title = "{Baryogenesis Without Grand Unification}",
    reportNumber = "RIFP-641",
    doi = "10.1016/0370-2693(86)91126-3",
    journal = "Phys. Lett. B",
    volume = "174",
    pages = "45--47",
    year = "1986"
}

@article{QRPA1,
  title = {Nuclear matrix elements for $0\ensuremath{\nu}\ensuremath{\beta}\ensuremath{\beta}$ decays with light or heavy Majorana-neutrino exchange},
  author = {Hyv\"arinen, Juhani and Suhonen, Jouni},
  journal = {Phys. Rev. C},
  volume = {91},
  issue = {2},
  pages = {024613},
  numpages = {12},
  year = {2015},
  month = {Feb},
  publisher = {American Physical Society},
  doi = {10.1103/PhysRevC.91.024613}
}

@article{EDF2,
  title = {Shape and Pairing Fluctuation Effects on Neutrinoless Double Beta Decay Nuclear Matrix Elements},
  author = {Vaquero, Nuria L\'opez and Rodr\'{\i}guez, Tom\'as R. and Egido, J. Luis},
  journal = {Phys. Rev. Lett.},
  volume = {111},
  issue = {14},
  pages = {142501},
  numpages = {5},
  year = {2013},
  month = {Sep},
  publisher = {American Physical Society},
  doi = {10.1103/PhysRevLett.111.142501}
}

@article{shell_model,
    author = "Coraggio, L. and Itaco, N. and De Gregorio, G. and Gargano, A. and Mancino, R. and Nowacki, F.",
    title = "{Shell-model calculation of $^{100}$Mo double-$\beta$ decay}",
    doi = "10.1103/PhysRevC.105.034312",
    journal = "Phys. Rev. C",
    volume = "105",
    number = "3",
    pages = "034312",
    year = "2022"
}

@article{PhysRevLett.120.202001,
  title = {New Leading Contribution to Neutrinoless Double-$\ensuremath{\beta}$ Decay},
  author = {Cirigliano, Vincenzo and Dekens, Wouter and de Vries, Jordy and Graesser, Michael L. and Mereghetti, Emanuele and Pastore, Saori and van Kolck, Ubirajara},
  journal = {Phys. Rev. Lett.},
  volume = {120},
  issue = {20},
  pages = {202001},
  numpages = {6},
  year = {2018},
  month = {May},
  publisher = {American Physical Society},
  doi = {10.1103/PhysRevLett.120.202001}
}

@article{PhysRevC.97.065501,
  title = {Neutrinoless double-$\ensuremath{\beta}$ decay in effective field theory: The light-Majorana neutrino-exchange mechanism},
  author = {Cirigliano, Vincenzo and Dekens, Wouter and Mereghetti, Emanuele and Walker-Loud, Andr\'e},
  journal = {Phys. Rev. C},
  volume = {97},
  issue = {6},
  pages = {065501},
  numpages = {13},
  year = {2018},
  month = {Jun},
  publisher = {American Physical Society},
  doi = {10.1103/PhysRevC.97.065501}
}

@article{PhysRevC.100.055504,
  title = {Renormalized approach to neutrinoless double-$\ensuremath{\beta}$ decay},
  author = {Cirigliano, V. and Dekens, W. and de Vries, J. and Graesser, M. L. and Mereghetti, E. and Pastore, S. and Piarulli, M. and van Kolck, U. and Wiringa, R. B.},
  journal = {Phys. Rev. C},
  volume = {100},
  issue = {5},
  pages = {055504},
  numpages = {36},
  year = {2019},
  month = {Nov},
  publisher = {American Physical Society},
  doi = {10.1103/PhysRevC.100.055504}
}

@article{PhysRevC.107.044305,
  title = {Neutrinoless $\ensuremath{\beta}\ensuremath{\beta}$-decay nuclear matrix elements from two-neutrino $\ensuremath{\beta}\ensuremath{\beta}$-decay data},
  author = {Jokiniemi, Lotta and Romeo, Beatriz and Soriano, Pablo and Men\'endez, Javier},
  journal = {Phys. Rev. C},
  volume = {107},
  issue = {4},
  pages = {044305},
  numpages = {11},
  year = {2023},
  month = {Apr},
  publisher = {American Physical Society},
  doi = {10.1103/PhysRevC.107.044305}
}

@inproceedings{10.1117/12.176771,
author = {Eugene E. Haller and K. M. Itoh and Jeffrey W. Beeman and William L. Hansen and V. I. Ozhogin},
title = {{Neutron transmutation doped natural and isotopically engineered germanium thermistors}},
volume = {2198},
booktitle = {Instrumentation in Astronomy VIII},
editor = {David L. Crawford and Eric R. Craine},
organization = {International Society for Optics and Photonics},
publisher = {SPIE},
pages = {630 -- 637},
year = {1994},
doi = {10.1117/12.176771}
}

@article{CUPID-Mo:2023lru,
    author = "{CUPID-Mo collaboration}",
    collaboration = "CUPID-Mo",
    title = "{Measurement of the 2\ensuremath{\nu}\ensuremath{\beta}\ensuremath{\beta} Decay Rate and Spectral Shape of Mo100 from the CUPID-Mo Experiment}",
    eprint = "2307.14086",
    archivePrefix = "arXiv",
    primaryClass = "nucl-ex",
    doi = "10.1103/PhysRevLett.131.162501",
    journal = "Phys. Rev. Lett.",
    volume = "131",
    number = "16",
    pages = "162501",
    year = "2023"
}

@article{Ahmine:2023xhg,
    author = "Ahmine, A. and others",
    title = "{Enhanced light signal for the suppression of pile-up events in Mo-based bolometers for the 0$\nu \beta \beta $ decay search.}",
    eprint = "2302.13944",
    archivePrefix = "arXiv",
    primaryClass = "physics.ins-det",
    doi = "10.1140/epjc/s10052-023-11519-6",
    journal = "Eur. Phys. J. C",
    volume = "83",
    number = "5",
    pages = "373",
    year = "2023"
}

@article{CUPID:2022opf,
    author = "{CUPID collaboration}",
    collaboration = "CUPID",
    title = "{Optimization of the first CUPID detector module}",
    eprint = "2202.06279",
    archivePrefix = "arXiv",
    primaryClass = "physics.ins-det",
    doi = "10.1140/epjc/s10052-022-10720-3",
    journal = "Eur. Phys. J. C",
    volume = "82",
    number = "9",
    pages = "810",
    year = "2022"
}

@article{CUPID_Baseline,
    author = "{CUPID collaboration}",
    collaboration = "CUPID",
    title = "{CUPID, the Cuore upgrade with particle identification}",
    eprint = "2503.02894",
    archivePrefix = "arXiv",
    primaryClass = "physics.ins-det",
    doi = "10.1140/epjc/s10052-025-14352-1",
    journal = "Eur. Phys. J. C",
    volume = "85",
    number = "7",
    pages = "737",
    year = "2025"
}

@article{CUPID_CDR,
      title="{CUPID pre-CDR}", 
      author="{CUPID collaboration}",
      year={2019},
      eprint={1907.09376},
      archivePrefix={arXiv},
      primaryClass = "physics.ins-det",
      journal = "CUPID pre-CDR, arXiv:1907.09376 [physics.ins-det]"
}

@article{CUORE_nature,
    author = "{CUORE collaboration}",
    collaboration = "CUORE",
    title = "{Search for Majorana neutrinos exploiting millikelvin cryogenics with CUORE}",
    eprint = "2104.06906",
    archivePrefix = "arXiv",
    primaryClass = "nucl-ex",
    doi = "10.1038/s41586-022-04497-4",
    journal = "Nature",
    volume = "604",
    number = "7904",
    pages = "53--58",
    year = "2022"
}

@article{discovery,
    author = "Ettengruber, Manuel and Agostini, Matteo and Caldwell, Allen and Eller, Philipp and Schulz, Oliver",
    title = "{Discovering neutrinoless double-beta decay in the era of precision neutrino cosmology}",
    eprint = "2208.09954",
    archivePrefix = "arXiv",
    primaryClass = "hep-ph",
    doi = "10.1103/PhysRevD.106.073004",
    journal = "Phys. Rev. D",
    volume = "106",
    number = "7",
    pages = "073004",
    year = "2022"
}

@article{NOVATI2019320,
title = {Charge-to-heat transducers exploiting the Neganov-Trofimov-Luke effect for light detection in rare-event searches},
journal = {Nucl. Instrum. Meth. Phys. Res. A},
volume = {940},
pages = {320-327},
year = {2019},
issn = {0168-9002},
doi = {https://doi.org/10.1016/j.nima.2019.06.044},
author = {Novati, V. and others}
}

@article{Cowan_freq,
    author = "Cowan, Glen and Cranmer, Kyle and Gross, Eilam and Vitells, Ofer",
    title = "{Asymptotic formulae for likelihood-based tests of new physics}",
    eprint = "1007.1727",
    archivePrefix = "arXiv",
    primaryClass = "physics.data-an",
    doi = "10.1140/epjc/s10052-011-1554-0",
    journal = "Eur. Phys. J. C",
    volume = "71",
    pages = "1554",
    year = "2011",
    note = "[Erratum: Eur.Phys.J.C 73, 2501 (2013)]"
}

@article{Feldman_cousins,
    author = "Feldman, Gary J. and Cousins, Robert D.",
    title = "{A Unified approach to the classical statistical analysis of small signals}",
    eprint = "physics/9711021",
    archivePrefix = "arXiv",
    reportNumber = "HUTP-97-A096",
    doi = "10.1103/PhysRevD.57.3873",
    journal = "Phys. Rev. D",
    volume = "57",
    pages = "3873--3889",
    year = "1998"
}

@article{rec_DM,
    author = "Baxter, D. and others",
    title = "{Recommended conventions for reporting results from direct dark matter searches}",
    eprint = "2105.00599",
    archivePrefix = "arXiv",
    primaryClass = "hep-ex",
    doi = "10.1140/epjc/s10052-021-09655-y",
    journal = "Eur. Phys. J. C",
    volume = "81",
    number = "10",
    pages = "907",
    year = "2021"
}

@article{Wilk,
    author = "Wilks, S. S.",
    title = "{The Large-Sample Distribution of the Likelihood Ratio for Testing Composite Hypotheses}",
    doi = "10.1214/aoms/1177732360",
    journal = "Annals Math. Statist.",
    volume = "9",
    number = "1",
    pages = "60--62",
    year = "1938"
}

@book{bayes,
  author = {Gelman, Andrew and Carlin, John B. and Stern, Hal S. and Rubin, Donald B.},
  edition = {2nd ed.},
  publisher = {Chapman and Hall/CRC},
  title = {Bayesian Data Analysis},
  year = 2004
}

@article{BAT,
author = {Caldwell, Allen and Koll{\'{a}}r, Daniel and Kr{\"{o}}ninger, Kevin},
journal = {Comput. Phys. Commun.},
month = {nov},
number = {11},
pages = {2197--2209},
publisher = {Elsevier B.V.},
volume = {180},
year = {2009},
title = {{BAT – The Bayesian analysis toolkit}},
}

@article{model_1,
    author = "Trotta, Roberto",
    title = "{Bayes in the sky: Bayesian inference and model selection in cosmology}",
    eprint = "0803.4089",
    archivePrefix = "arXiv",
    primaryClass = "astro-ph",
    doi = "10.1080/00107510802066753",
    journal = "Contemp. Phys.",
    volume = "49",
    pages = "71--104",
    year = "2008"
}

@article{model_2,
    author = "Caldwell, Allen and Kroninger, Kevin",
    title = "{Signal discovery in sparse spectra: A Bayesian analysis}",
    eprint = "physics/0608249",
    archivePrefix = "arXiv",
    reportNumber = "MPP-2006-107",
    doi = "10.1103/PhysRevD.74.092003",
    journal = "Phys. Rev. D",
    volume = "74",
    pages = "092003",
    year = "2006"
}

@article{GERDA:2020xhi,
    author = "Agostini, M. and others",
    collaboration = "GERDA",
    title = "{Final Results of GERDA on the Search for Neutrinoless Double-$\beta$ Decay}",
    eprint = "2009.06079",
    archivePrefix = "arXiv",
    primaryClass = "nucl-ex",
    doi = "10.1103/PhysRevLett.125.252502",
    journal = "Phys. Rev. Lett.",
    volume = "125",
    number = "25",
    pages = "252502",
    year = "2020"
}

@article{kot_psf,
    author = "Kotila, J. and Iachello, F.",
    title = "{Phase space factors for double-$\beta$ decay}",
    eprint = "1209.5722",
    archivePrefix = "arXiv",
    primaryClass = "nucl-th",
    doi = "10.1103/PhysRevC.85.034316",
    journal = "Phys. Rev. C",
    volume = "85",
    pages = "034316",
    year = "2012"
}

@article{Barabash:2b2n,
    author = "Barabash, Alexander",
    title = "{Precise Half-Life Values for Two-Neutrino Double-\ensuremath{\beta} Decay: 2020 Review}",
    eprint = "2009.14451",
    archivePrefix = "arXiv",
    primaryClass = "nucl-ex",
    doi = "10.3390/universe6100159",
    journal = "Universe",
    volume = "6",
    number = "10",
    pages = "159",
    year = "2020"
}

@article{Agostini:2022zub,
    author = "Agostini, Matteo and Benato, Giovanni and Detwiler, Jason A. and Men\'endez, Javier and Vissani, Francesco",
    title = "{Toward the discovery of matter creation with neutrinoless \ensuremath{\beta}\ensuremath{\beta} decay}",
    eprint = "2202.01787",
    archivePrefix = "arXiv",
    primaryClass = "hep-ex",
    doi = "10.1103/RevModPhys.95.025002",
    journal = "Rev. Mod. Phys.",
    volume = "95",
    number = "2",
    pages = "025002",
    year = "2023"
}

@article{CUORE:2002myo,
    author = "{CUORE collaboration}",
    collaboration = "CUORE",
    title = "{CUORE: A Cryogenic underground observatory for rare events}",
    eprint = "hep-ex/0212053",
    archivePrefix = "arXiv",
    doi = "10.1016/j.nima.2003.07.067",
    journal = "Nucl. Instrum. Meth. A",
    volume = "518",
    pages = "775--798",
    year = "2004"
}

@article{CUORE:2015hsf,
    author = "{CUORE collaboration}",
    collaboration = "CUORE",
    title = "{Search for Neutrinoless Double-Beta Decay of $^{130}$Te with CUORE-0}",
    eprint = "1504.02454",
    archivePrefix = "arXiv",
    primaryClass = "nucl-ex",
    doi = "10.1103/PhysRevLett.115.102502",
    journal = "Phys. Rev. Lett.",
    volume = "115",
    number = "10",
    pages = "102502",
    year = "2015"
}

@article{CUORE:2021ctv,
    author = "{CUORE collaboration}",
    collaboration = "CUORE",
    title = "{CUORE opens the door to tonne-scale cryogenics experiments}",
    eprint = "2108.07883",
    archivePrefix = "arXiv",
    primaryClass = "physics.ins-det",
    doi = "10.1016/j.ppnp.2021.103902",
    journal = "Prog. Part. Nucl. Phys.",
    volume = "122",
    pages = "103902",
    year = "2022"
}

@article{Poda:2021hsv,
    author = "Poda, Denys",
    title = "{Scintillation in Low-Temperature Particle Detectors}",
    doi = "10.3390/physics3030032",
    journal = "MDPI Physics",
    volume = "3",
    number = "3",
    pages = "473--535",
    year = "2021"
}

@article{Armengaud:2019loe,
    author = "{CUPID-Mo collaboration}",
    title = "{The CUPID-Mo experiment for neutrinoless double-beta decay: performance and prospects}",
    eprint = "1909.02994",
    archivePrefix = "arXiv",
    primaryClass = "physics.ins-det",
    doi = "10.1140/epjc/s10052-019-7578-6",
    journal = "Eur. Phys. J. C",
    volume = "80",
    number = "1",
    pages = "44",
    year = "2020"
}

@article{Augier:2022znx,
    author = "{CUPID-Mo collaboration}",
    title = "{Final results on the $0\nu \beta \beta $ decay half-life limit of $^{100}$Mo from the CUPID-Mo experiment}",
    eprint = "2202.08716",
    archivePrefix = "arXiv",
    primaryClass = "nucl-ex",
    doi = "10.1140/epjc/s10052-022-10942-5",
    journal = "Eur. Phys. J. C",
    volume = "82",
    number = "11",
    pages = "1033",
    year = "2022"
}

@article{CUPID:2022puj,
    author = "{CUPID-0 collaboration}",
    collaboration = "CUPID-0",
    title = "{Final Result on the Neutrinoless Double Beta Decay of $^{82}$Se with CUPID-0}",
    eprint = "2206.05130",
    archivePrefix = "arXiv",
    primaryClass = "hep-ex",
    doi = "10.1103/PhysRevLett.129.111801",
    journal = "Phys. Rev. Lett.",
    volume = "129",
    number = "11",
    pages = "111801",
    year = "2022"
}

@article{CUPID:2019lzs,
    author = "{CUPID-0 collaboration}",
    collaboration = "CUPID-0",
    title = "{Background Model of the CUPID-0 Experiment}",
    eprint = "1904.10397",
    archivePrefix = "arXiv",
    primaryClass = "nucl-ex",
    doi = "10.1140/epjc/s10052-019-7078-8",
    journal = "Eur. Phys. J. C",
    volume = "79",
    number = "7",
    pages = "583",
    year = "2019"
}

@article{Deppisch:2017ecm,
    author = "Deppisch, Frank F. and Graf, Lukas and Harz, Julia and Huang, Wei-Chih",
    title = "{Neutrinoless Double Beta Decay and the Baryon Asymmetry of the Universe}",
    eprint = "1711.10432",
    archivePrefix = "arXiv",
    primaryClass = "hep-ph",
    reportNumber = "DO-TH-17-19, CP3-ORIGINS-2017-055, DO-TH 17/19, CP3-Origins-2017-055 DNRF90",
    doi = "10.1103/PhysRevD.98.055029",
    journal = "Phys. Rev. D",
    volume = "98",
    number = "5",
    pages = "055029",
    year = "2018"
}

@article{Dolinski:2019nrj,
    author = "Dolinski, Michelle J. and Poon, Alan W. P. and Rodejohann, Werner",
    title = "{Neutrinoless Double-Beta Decay: Status and Prospects}",
    eprint = "1902.04097",
    archivePrefix = "arXiv",
    primaryClass = "nucl-ex",
    doi = "10.1146/annurev-nucl-101918-023407",
    journal = "Ann. Rev. Nucl. Part. Sci.",
    volume = "69",
    pages = "219--251",
    year = "2019"
}

@article{DellOro:2016tmg,
    author = "Dell'Oro, Stefano and Marcocci, Simone and Viel, Matteo and Vissani, Francesco",
    title = "{Neutrinoless double beta decay: 2015 review}",
    eprint = "1601.07512",
    archivePrefix = "arXiv",
    primaryClass = "hep-ph",
    doi = "10.1155/2016/2162659",
    journal = "Adv. High Energy Phys.",
    volume = "2016",
    pages = "2162659",
    year = "2016"
}

@article{CUORE:2024ikf,
    author = "{CUORE collaboration}",
    collaboration = "CUORE",
    title = "{Constraints on Lepton Number Violation with the 2 tonne$\cdot$yr CUORE Dataset}",
    journal = {Science},
    volume = {390},
    number = {6777},
    pages = {1029-1032},
    year = {2025},
    eprint = "2404.04453",
    archivePrefix = "arXiv",
    primaryClass = "nucl-ex",
    doi = "10.1126/science.adp6474"
}

@article{Luke:1988,
  author="Luke, {\relax P. N}",
  title="Voltage-assisted calorimetric ionization detector",
  journal="J. Appl. Phys.",
  volume="64",
  pages="6858",
  year="1988"
  }

@article{Neganov:1985khw,
    author = "Neganov, B. S. and Trofimov, V. N.",
    title = "{Colorimetric method measuring ionizing radiation}",
    reportNumber = "USSR Patent No 1037771",
    journal = "Otkryt. Izobret.",
    volume = "146",
    pages = "215",
    year = "1985"
}

@misc{Cowan,
      title={Statistical Methods for Particle Physics
Lecture 3: parameter estimation}, 
      author={Glen Cowan},
      year={2014},
      url={www.pp.rhul.ac.uk/~cowan/stat_aachen.html}, 
}

@article{CUPID:2018kff,
    author = "{CUPID-0 collaboration}",
    collaboration = "CUPID-0",
    title = "{CUPID-0: the first array of enriched scintillating bolometers for $0\nu\beta\beta$ decay investigations}",
    eprint = "1802.06562",
    archivePrefix = "arXiv",
    primaryClass = "physics.ins-det",
    doi = "10.1140/epjc/s10052-018-5896-8",
    journal = "Eur. Phys. J. C",
    volume = "78",
    number = "5",
    pages = "428",
    year = "2018"
}

@article{Pritychenko:2024ase,
    author = "Pritychenko, B. and Tretyak, V. I.",
    title = "{Comprehensive review of 2\ensuremath{\beta} decay half-lives}",
    doi = "10.1016/j.adt.2024.101694",
    journal = "Atom. Data Nucl. Data Tabl.",
    volume = "161",
    pages = "101694",
    year = "2025"
}

@article{Gomez-Cadenas:2023vca,
    author = "G\'omez-Cadenas, Juan Jos\'e and Mart\'\i{}n-Albo, Justo and Men\'endez, Javier and Mezzetto, Mauro and Monrabal, Francesc and Sorel, Michel",
    title = "{The search for neutrinoless double-beta decay}",
    doi = "10.1007/s40766-023-00049-2",
    journal = "Riv. Nuovo Cim.",
    volume = "46",
    number = "10",
    pages = "619--692",
    year = "2023"
}

@article{Bossio:2023wpj,
    author = "Bossio, Elisabetta and Agostini, Matteo",
    title = "{Probing beyond the standard model physics with double-beta decays}",
    eprint = "2304.07198",
    archivePrefix = "arXiv",
    primaryClass = "hep-ex",
    doi = "10.1088/1361-6471/ad11f9",
    journal = "J. Phys. G",
    volume = "51",
    number = "2",
    pages = "023001",
    year = "2024"
}

@article{Gysbers:2019uyb,
    author = "Gysbers, P. and others",
    title = "{Discrepancy between experimental and theoretical $\beta$-decay rates resolved from first principles}",
    eprint = "1903.00047",
    archivePrefix = "arXiv",
    primaryClass = "nucl-th",
    doi = "10.1038/s41567-019-0450-7",
    journal = "Nature Phys.",
    volume = "15",
    number = "5",
    pages = "428--431",
    year = "2019"
}

@article{Suhonen:2017krv,
    author = "Suhonen, Jouni T.",
    title = "{Value of the Axial-Vector Coupling Strength in {\ensuremath{\beta}} and {\ensuremath{\beta}}{\ensuremath{\beta}} Decays: A Review}",
    eprint = "1712.01565",
    archivePrefix = "arXiv",
    primaryClass = "nucl-th",
    doi = "10.3389/fphy.2017.00055",
    journal = "Front. in Phys.",
    volume = "5",
    pages = "55",
    year = "2017"
}

@phdthesis{loaiza:tel-04633827,
  TITLE = {{Background studies in neutrinoless double beta decay and dark matter searches}},
  AUTHOR = {Loaiza, Pia},
  URL = {https://hal.science/tel-04633827},
  SCHOOL = {{Universit{\'e} Paris-Saclay}},
  YEAR = {2024},
  MONTH = Mar,
  TYPE = {Habilitation {\`a} diriger des recherches},
  HAL_ID = {tel-04633827},
  HAL_VERSION = {v1},
}

@phdthesis{imbert:tel-04266831,
  TITLE = {{{\'E}tude du bruit de fond des exp{\'e}riences CUPID-Mo, CUPID, et CROSS  de double d{\'e}sint{\'e}gration b{\^e}ta sans {\'e}mission de neutrinos}},
  AUTHOR = {Imbert, L{\'e}onard},
  URL = {https://theses.hal.science/tel-04266831},
  NUMBER = {2023UPASP086},
  SCHOOL = {{Universit{\'e} Paris-Saclay}},
  YEAR = {2023},
  MONTH = Sep,
  TYPE = {Theses},
  HAL_ID = {tel-04266831},
  HAL_VERSION = {v1},
}

@article{Suhonen:2013laa,
    author = "Suhonen, Jouni and Civitarese, Osvaldo",
    title = "{Probing the quenching of g$_{A}$ by single and double beta decays}",
    doi = "10.1016/j.physletb.2013.06.042",
    journal = "Phys. Lett. B",
    volume = "725",
    pages = "153--157",
    year = "2013"
}

@article{Capozzi:2025wyn,
    author = "Capozzi, Francesco and Giar{\`e}, William and Lisi, Eligio and Marrone, Antonio and Melchiorri, Alessandro and Palazzo, Antonio",
    title = "{Neutrino masses and mixing: Entering the era of subpercent precision}",
    eprint = "2503.07752",
    archivePrefix = "arXiv",
    primaryClass = "hep-ph",
    doi = "10.1103/PhysRevD.111.093006",
    journal = "Phys. Rev. D",
    volume = "111",
    number = "9",
    pages = "093006",
    year = "2025"
}

@article{ParticleDataGroup:2024cfk,
    author = "Navas, S. and others",
    collaboration = "Particle Data Group",
    title = "{Review of particle physics}",
    doi = "10.1103/PhysRevD.110.030001",
    journal = "Phys. Rev. D",
    volume = "110",
    number = "3",
    pages = "030001",
    year = "2024"
}

@article{CUORE:2024fak,
    author = "{CUORE collaboration}",
    collaboration = "CUORE",
    title = "{Data-driven background model for the CUORE experiment}",
    eprint = "2405.17937",
    archivePrefix = "arXiv",
    primaryClass = "nucl-ex",
    doi = "10.1103/PhysRevD.110.052003",
    journal = "Phys. Rev. D",
    volume = "110",
    number = "5",
    pages = "052003",
    year = "2024"
}

@article{CUPID-Mo:2023vle,
    author = "{CUPID-Mo collaboration}",
    collaboration = "CUPID-Mo",
    title = "{The background model of the CUPID-Mo $0\nu \beta \beta $ experiment}",
    eprint = "2305.01402",
    archivePrefix = "arXiv",
    primaryClass = "hep-ex",
    doi = "10.1140/epjc/s10052-023-11830-2",
    journal = "Eur. Phys. J. C",
    volume = "83",
    number = "7",
    pages = "675",
    year = "2023"
}

@article{CUPID:2025wbt,
    author = "{CUPID collaboration}",
    collaboration = "CUPID",
    title = "{A gravity-based mounting approach for large-scale cryogenic calorimeter arrays}",
    doi = "10.1140/epjc/s10052-025-14613-z",
    journal = "Eur. Phys. J. C",
    volume = "85",
    number = "9",
    pages = "935",
    year = "2025"
}

@article{25tk-nctn,
  title = {First results on the search for lepton number violating neutrinoless double-$\ensuremath{\beta}$ decay with the LEGEND-200 experiment},
  author = "{LEGEND collaboration}",
  journal = {Phys. Rev. Lett.},
  year = {2025},
  month = {Sep},
  publisher = {American Physical Society},
  doi = {10.1103/25tk-nctn}
}

@article{KamLAND-Zen:2024eml,
    author = "{KamLAND-Zen collaboration}",
    collaboration = "KamLAND-Zen",
    title = "{Search for Majorana Neutrinos with the Complete KamLAND-Zen Dataset}",
    eprint = "2406.11438",
    archivePrefix = "arXiv",
    primaryClass = "hep-ex",
    month = "6",
    year = "2024",
    journal = "arXiv:2406.11438 [hep-ex]"
}

@article{LEGEND:2021bnm,
    author = "{LEGEND collaboration}",
    collaboration = "LEGEND",
    title = "{The Large Enriched Germanium Experiment for Neutrinoless $\beta\beta$ Decay}: {LEGEND-1000 Preconceptual Design Report}",
    eprint = "2107.11462",
    archivePrefix = "arXiv",
    primaryClass = "physics.ins-det",
    month = "7",
    year = "2021",
    journal = "arXiv:2107.11462 [physics.ins-det]"
}

@article{nEXO:2021ujk,
    author = "{nEXO collaboration}",
    collaboration = "nEXO",
    title = "{nEXO: neutrinoless double beta decay search beyond 10$^{28}$ year half-life sensitivity}",
    eprint = "2106.16243",
    archivePrefix = "arXiv",
    primaryClass = "nucl-ex",
    doi = "10.1088/1361-6471/ac3631",
    journal = "J. Phys. G",
    volume = "49",
    number = "1",
    pages = "015104",
    year = "2022"
}

@article{NEXT:2020amj,
    author = "{NEXT collaboration}",
    collaboration = "NEXT",
    title = "{Sensitivity of a tonne-scale NEXT detector for neutrinoless double beta decay searches}",
    eprint = "2005.06467",
    archivePrefix = "arXiv",
    primaryClass = "physics.ins-det",
    doi = "10.1007/JHEP08(2021)164",
    journal = "JHEP",
    volume = "2021",
    number = "08",
    pages = "164",
    year = "2021"
}

@misc{KamLand2Zen,
  author       = {Kelly Weerman and others},
  title        = "{Latest KamLAND-Zen results and the impact of muon spallation on the 0$\nu\beta\beta$ search}",
  note         = "{Talk at TAUP 2025}" ,
  month        = aug,
  year         = 2025
}

@article{Armatol:2025zrq,
    author = "Armatol, A. and others",
    title = "{Cryogenic light detectors with thermal signal amplification for $0\nu\beta\beta$ search experiments}",
    eprint = "2507.15732",
    archivePrefix = "arXiv",
    primaryClass = "physics.ins-det",
    month = "7",
    year = "2025",
    journal = "arXiv:2507.15732 [physics.ins-det]"
}

@article{CUORE:2025lzd,
    author = "{CUORE collaboration}",
    collaboration = "CUORE",
    title = "{Reconstruction of cosmic-ray muon events with CUORE}",
    eprint = "2509.05528",
    archivePrefix = "arXiv",
    primaryClass = "physics.ins-det",
    month = "9",
    year = "2025",
    journal = "arXiv:2509.05528 [physics.ins-det]"
}

@article{Moore:2025eil,
    author = "Moore, Maya and others",
    title = "{Performance of a SiPM-based, plastic scintillator muon veto prototype for CUPID}",
    eprint = "2505.06129",
    archivePrefix = "arXiv",
    primaryClass = "physics.ins-det",
    doi = "10.1088/1748-0221/20/08/P08020",
    journal = "JINST",
    volume = "20",
    number = "08",
    pages = "P08020",
    year = "2025"
}

@article{Simkovic:2018hiq,
    author = "{\v{S}}imkovic, Fedor and Smetana, Adam and Vogel, Petr",
    title = "{$0\nu\beta\beta$ nuclear matrix elements, neutrino potentials and $\mathrm{SU}(4)$ symmetry}",
    eprint = "1808.05016",
    archivePrefix = "arXiv",
    primaryClass = "nucl-th",
    doi = "10.1103/PhysRevC.98.064325",
    journal = "Phys. Rev. C",
    volume = "98",
    number = "6",
    pages = "064325",
    year = "2018"
}

@article{Deppisch:2020ztt,
    author = "Deppisch, Frank F. and Graf, Lukas and Iachello, Francesco and Kotila, Jenni",
    title = "{Analysis of light neutrino exchange and short-range mechanisms in $0\nu\beta\beta$ decay}",
    eprint = "2009.10119",
    archivePrefix = "arXiv",
    primaryClass = "hep-ph",
    doi = "10.1103/PhysRevD.102.095016",
    journal = "Phys. Rev. D",
    volume = "102",
    number = "9",
    pages = "095016",
    year = "2020"
}

@article{Rodriguez:2010mn,
    author = "Rodriguez, Tomas R. and Martinez-Pinedo, G.",
    title = "{Energy density functional study of nuclear matrix elements for neutrinoless $\beta\beta$ decay}",
    eprint = "1008.5260",
    archivePrefix = "arXiv",
    primaryClass = "nucl-th",
    doi = "10.1103/PhysRevLett.105.252503",
    journal = "Phys. Rev. Lett.",
    volume = "105",
    pages = "252503",
    year = "2010"
}

@article{Song:2017ktj,
    author = "Song, L. S. and Yao, J. M. and Ring, P. and Meng, J.",
    title = "{Nuclear matrix element of neutrinoless double-$\beta$ decay: Relativity and short-range correlations}",
    eprint = "1702.02448",
    archivePrefix = "arXiv",
    primaryClass = "nucl-th",
    doi = "10.1103/PhysRevC.95.024305",
    journal = "Phys. Rev. C",
    volume = "95",
    number = "2",
    pages = "024305",
    year = "2017"
}

@article{Vetter:2023fas,
    author = "Vetter, Kenneth J. and others",
    title = "{Improving the performance of cryogenic calorimeters with nonlinear multivariate noise cancellation algorithms}",
    eprint = "2311.01131",
    archivePrefix = "arXiv",
    primaryClass = "physics.ins-det",
    doi = "10.1140/epjc/s10052-024-12595-y",
    journal = "Eur. Phys. J. C",
    volume = "84",
    number = "3",
    pages = "243",
    year = "2024"
}

@article{Castillo:2024jfj,
    author = "Castillo, Daniel and Jokiniemi, Lotta and Soriano, Pablo and Men{\'e}ndez, Javier",
    title = "{Neutrinoless {\ensuremath{\beta}}{\ensuremath{\beta}} decay nuclear matrix elements complete up to N2LO in heavy nuclei}",
    eprint = "2408.03373",
    archivePrefix = "arXiv",
    primaryClass = "nucl-th",
    reportNumber = "INT-PUB-24-037",
    doi = "10.1016/j.physletb.2025.139851",
    journal = "Phys. Lett. B",
    volume = "860",
    pages = "139181",
    year = "2025",
    note = "[Erratum: Phys.Lett.B 869, 139851 (2025)]"
}

@article{Kauppinen:2025odt,
    author = "Kauppinen, Elina and Kotila, Jenni",
    title = "{Leading-order short-range nuclear matrix elements in double-{\ensuremath{\beta}} decay using the microscopic interacting boson model}",
    doi = "10.1103/6y3v-5fww",
    journal = "Phys. Rev. C",
    volume = "112",
    number = "3",
    pages = "034329",
    year = "2025"
}

@article{CUPID-Mo:2022cel,
    author = "{CUPID-Mo collaboration}",
    collaboration = "CUPID-Mo",
    title = "{New measurement of double-{\ensuremath{\beta}} decays of $^{100}$Mo to excited states of $^{100}$Ru with the CUPID-Mo experiment}",
    eprint = "2207.09577",
    archivePrefix = "arXiv",
    primaryClass = "nucl-ex",
    doi = "10.1103/PhysRevC.107.025503",
    journal = "Phys. Rev. C",
    volume = "107",
    number = "2",
    pages = "025503",
    year = "2023"
}
\section*{Appendix}

\begin{figure*}[h!]
    \centering
    \includegraphics[width=0.4\textwidth]{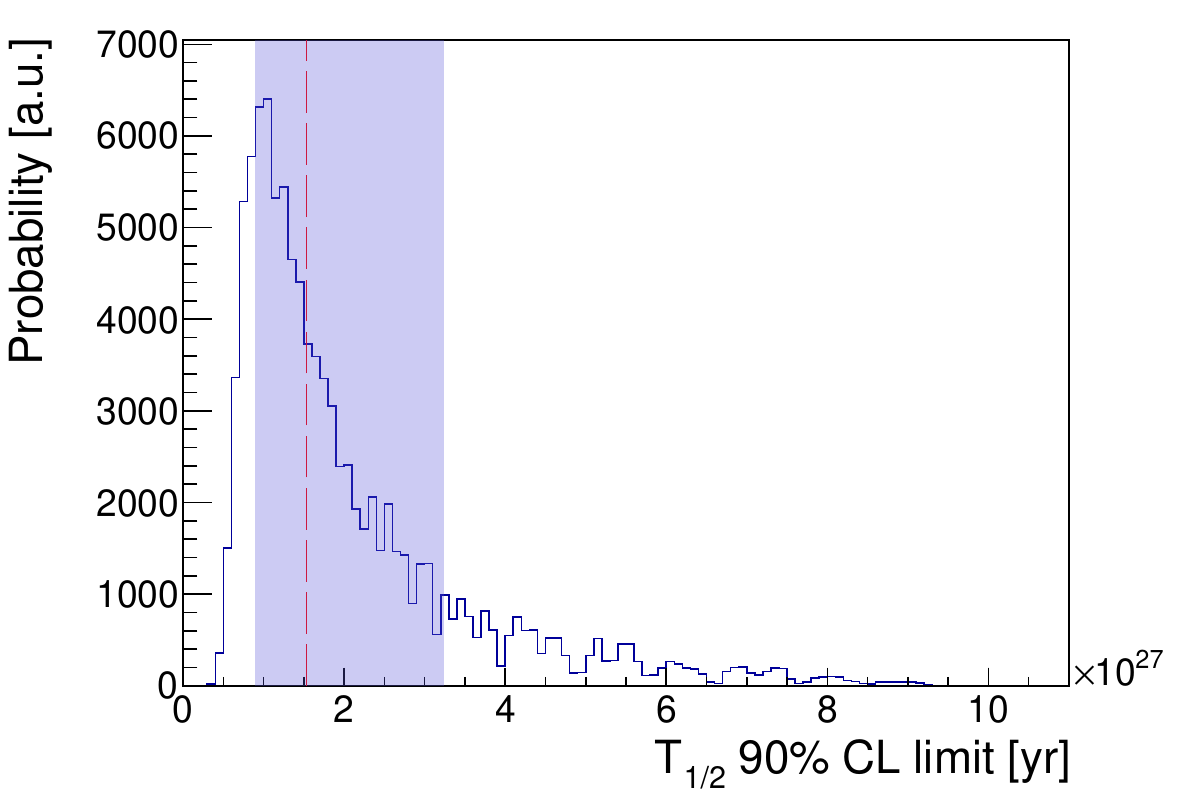}
       \includegraphics[width=0.4\textwidth]{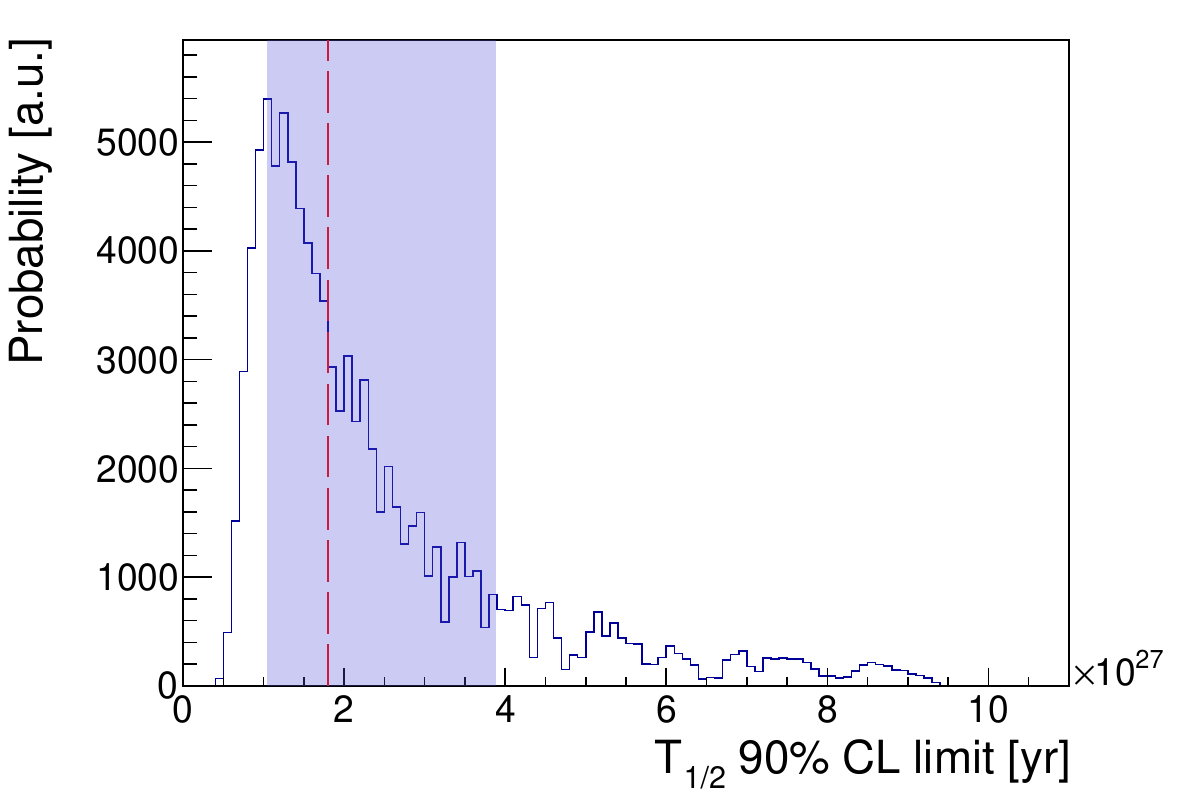}
    \includegraphics[width=0.4\textwidth]{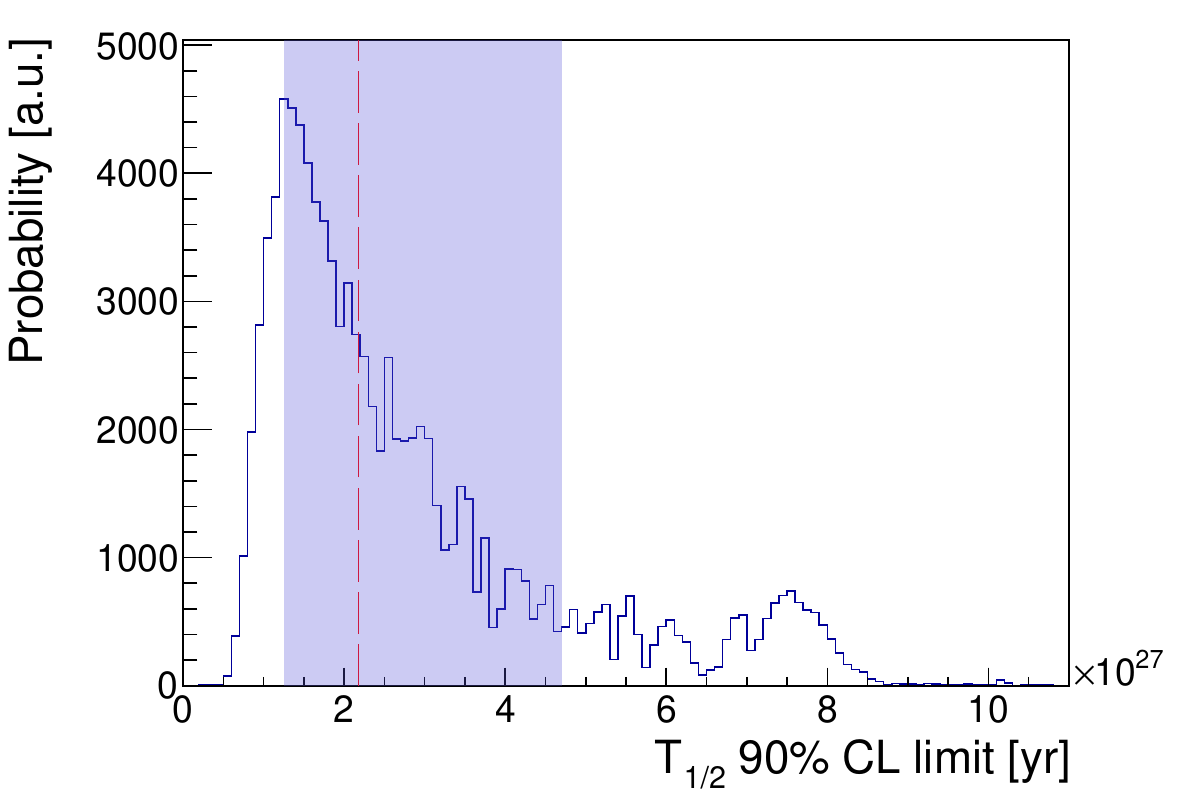}
    \includegraphics[width=0.4\textwidth]{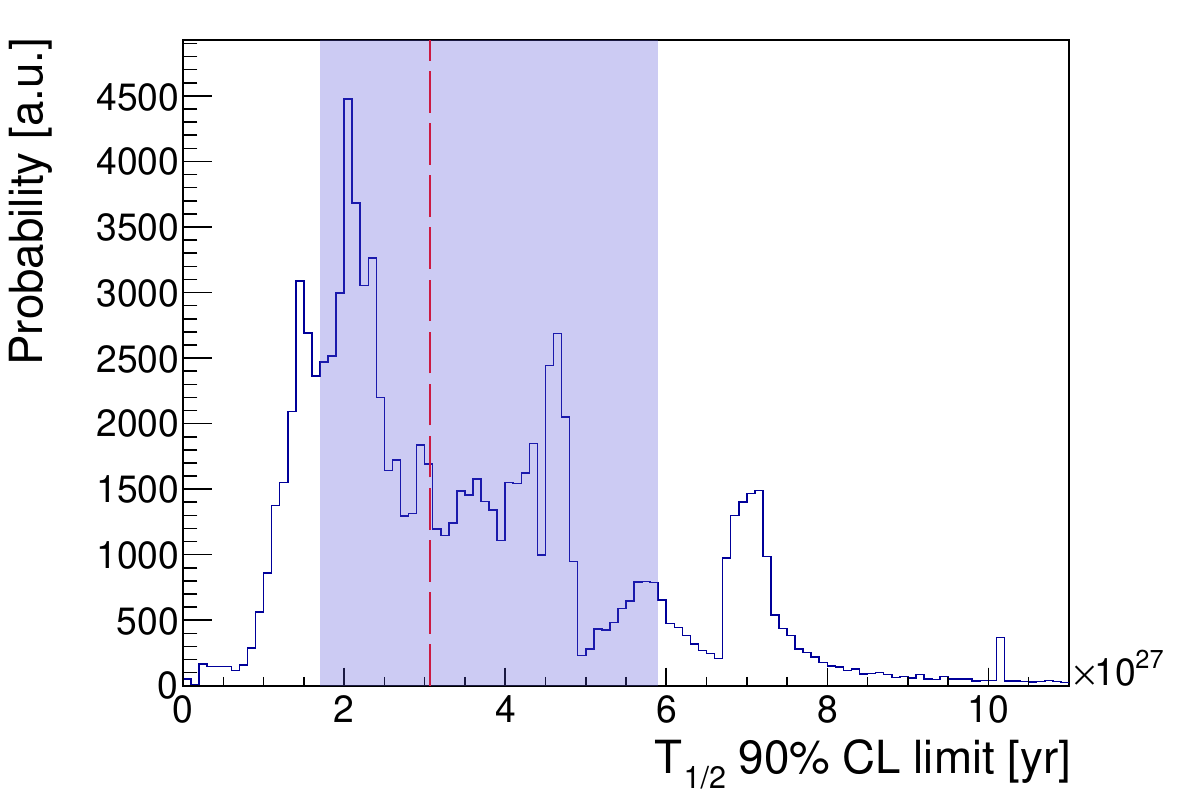}
    \caption{Distribution of 90\% CL exclusion limits obtained with a Frequentist analysis for CUPID under 4 background hypotheses  of $1.5~\times~10^{-4}$ counts/keV/kg/yr, $1.0~\times~10^{-4}$ counts/keV/kg/yr, 0.6~$\times~10^{-4}$ counts/keV/kg/yr  and 0.2~$\times ~10^{-4}$ counts/keV/kg/yr  from top left to bottom right. The median sensitivity and the interval containing 68\% of the pseudo-experiments are shown. The discretisation present in the plots with lower background index corresponds to discrete counts around the Q$_{\beta\beta}$.}
    \label{freq_distributions}
\end{figure*}

The supplementary material in this section provides further details on the results on exclusion sensitivities.
Figure \ref{freq_distributions} presents the   90\% CL exclusion sensitivity distribution, in the Frequentist framework, assuming the background levels of $1.5~\times~10^{-4}$ counts/keV/kg/yr, $1.0~\times~10^{-4}$ counts/keV/kg/yr, 0.6~$\times~10^{-4}$ counts/keV/kg/yr  and 0.2~$\times ~10^{-4}$ counts/keV/kg/yr. From these  distributions we obtain the median exclusion sensitivity  given in Table \ref{tab:exclusion}. For the ultra-low background index of 0.2  $\times$ 10$^{-4}$ counts/keV/kg/yr we get very few events. This is reflected in the discretization seen in  the probability distribution. Thus, it is worthy to note that the most probable results can be significantly different from the reported median sensitivity.

\begin{figure*}[h!]
    \centering
    \includegraphics[width=0.4\textwidth]{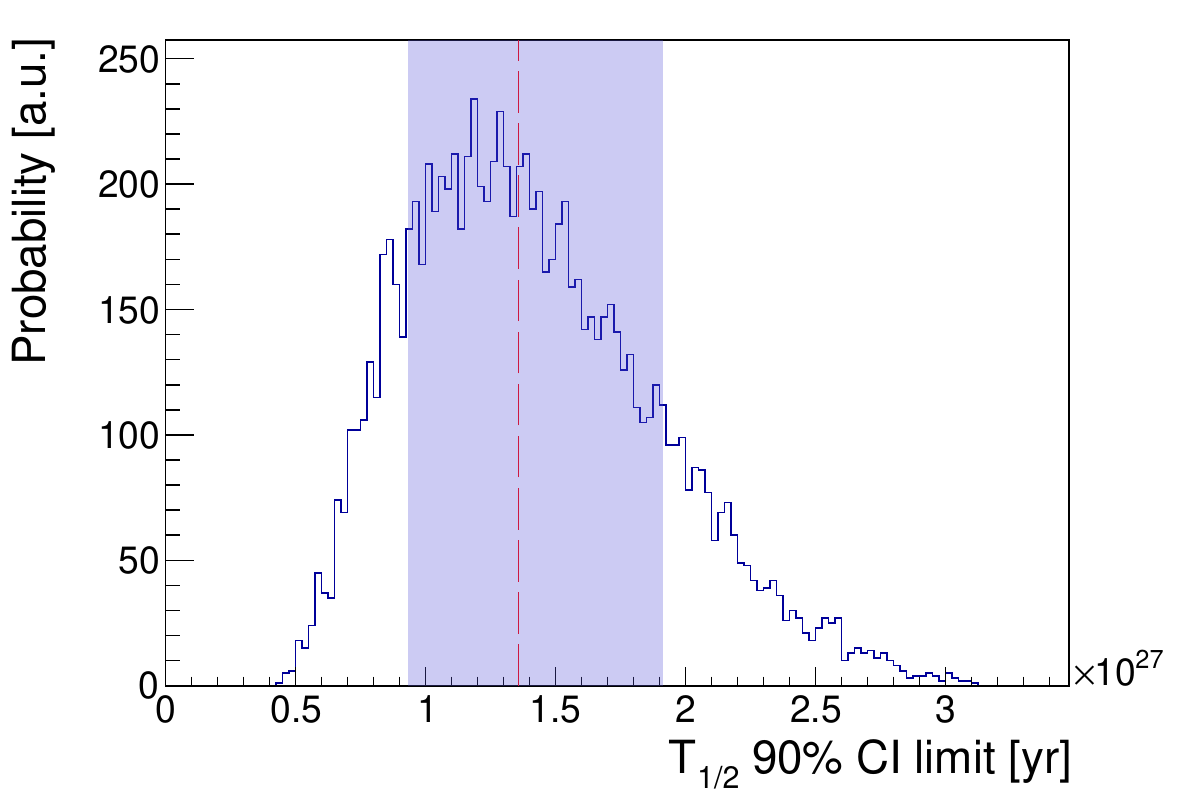}
    \includegraphics[width=0.4\textwidth]{figures/exclusion_limits_distribution_bay_base.pdf}
    \includegraphics[width=0.4\textwidth]{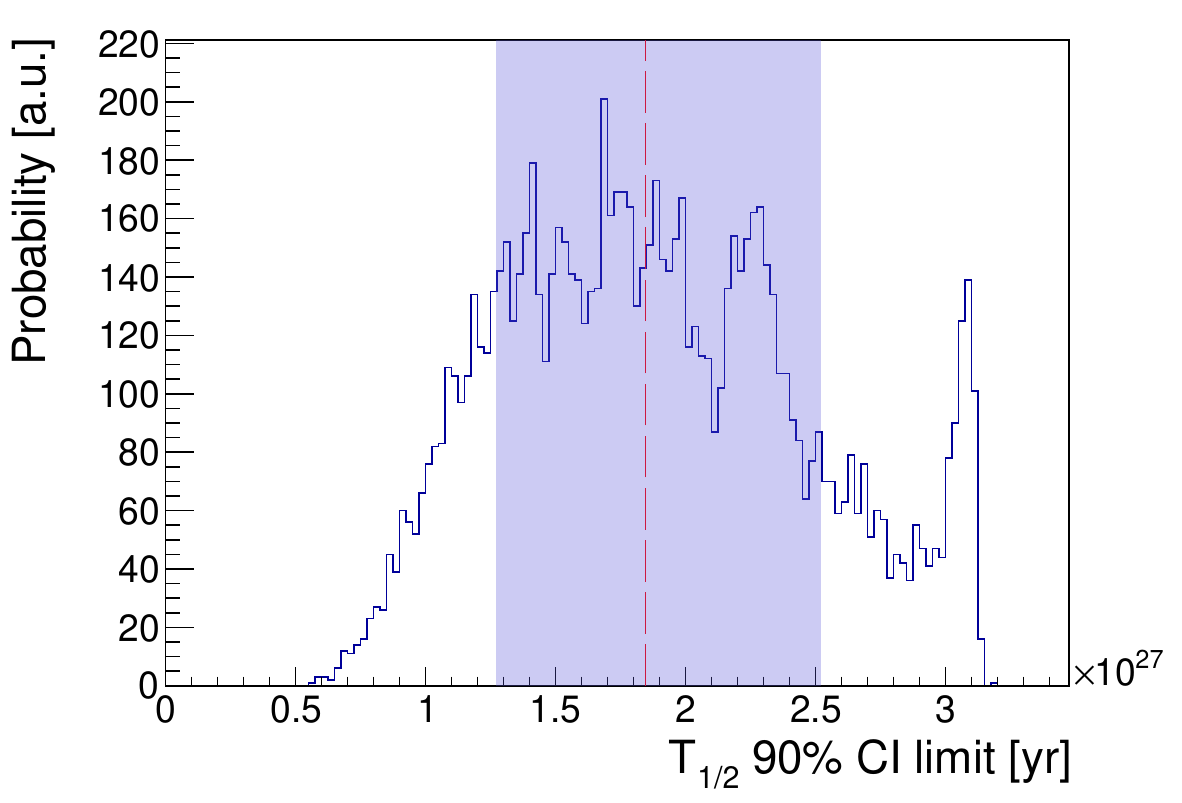}
    \includegraphics[width=0.4\textwidth]{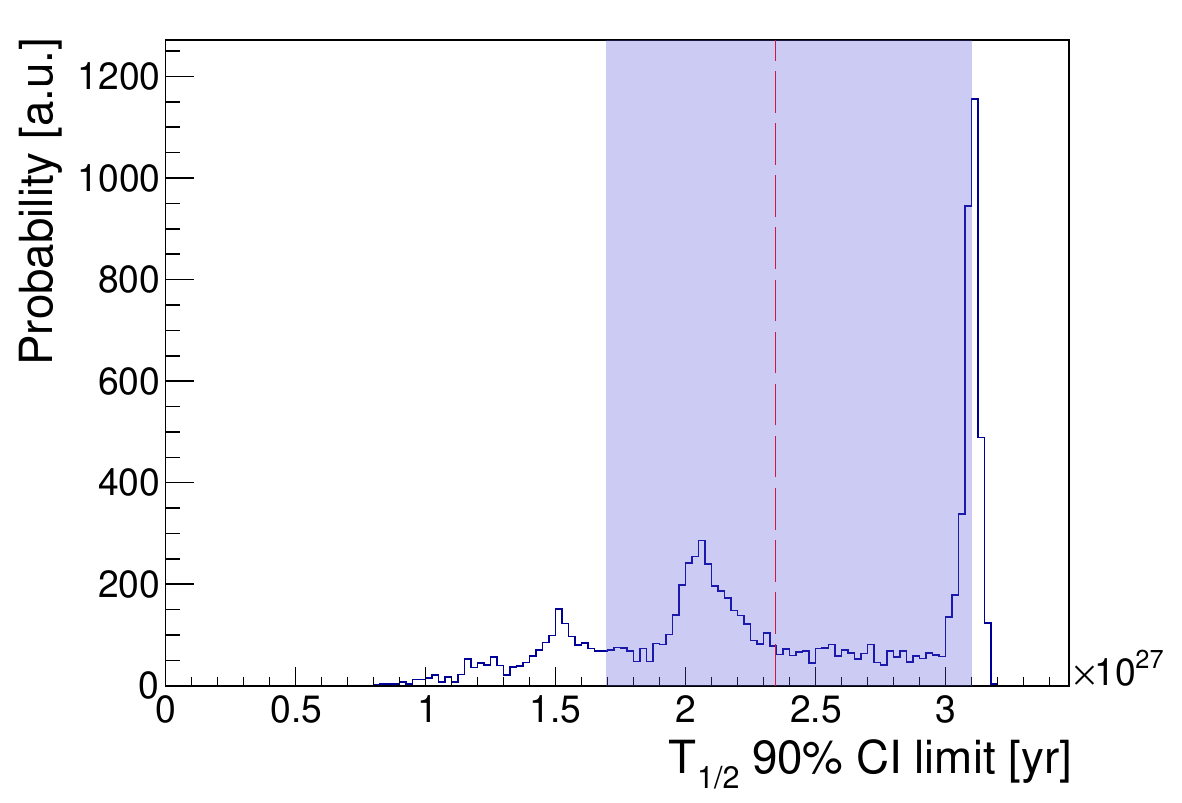}
    \caption{Distribution of 90\% c.i. exclusion limits for CUPID obtained with a Bayesian analysis, under the assumption of a background of  of $1.5~\times~10^{-4}$ counts/keV/kg/yr, $1.0~\times~10^{-4}$ counts/keV/kg/yr, 0.6~$\times~10^{-4}$ counts/keV/kg/yr  and 0.2~$\times ~10^{-4}$ counts/keV/kg/yr (top left, top right, bottom left, bottom right). The median sensitivity and interval containing 68\% c.i. of the pseudo-experiments are also shown. Peaks are visible in the case of a lower background corresponding to discrete counts around the Q$_{\beta\beta}$.}
    \label{bay_dist_tau}
\end{figure*}

Similarly, we show in Fig. \ref{bay_dist_tau} the distributions  of the 90\% c.i. exclusion sensitivities in the Bayesian framework for the four different background levels, from which we obtain the median exclusion sensitivities in Table \ref{tab:exclusion}. Similarly to the Frequentist analysis, peaks are visible in the distributions corresponding to low background index corresponding to a discrete number of counts around the Q$_{\beta\beta}$. 



\end{document}